\documentclass{aa}
\usepackage{graphicx}
\usepackage{natbib}
\usepackage{html}
\usepackage{txfonts}
\usepackage{deluxetable} 

\begin{document}
 \title{HST/STIS High Resolution Echelle Spectra\\ of \object{$\alpha$ Centauri
A} (G2
V)\footnote{Based on observations made with the NASA/ESA Hubble Space
Telescope,  obtained at the Space Telescope Science Institute, which is
operated by the  Association of Universities for Research in Astronomy, Inc.,
under NASA contract  NAS 5-26555. These observations are associated with
proposal GO-07263.}}

   \subtitle{ }

   \author{Isabella Pagano
          \inst{1}
          \and
          Jeffrey L. Linsky\inst{2}
\and 
Jeff Valenti\inst{3}
\and 
Douglas K. Duncan\inst{4}}

   \offprints{I. Pagano}

   \institute{INAF, Catania Astrophysical Observatory, via Santa
 Sofia 78, 95125 Catania, Italy\\
              \email{ipa@ct.astro.it}
         \and
             JILA, University of Colorado and NIST, Boulder, CO 80309-0440,
USA\\
             \email{jlinsky@jila.colorado.edu}
         \and
             Space Telescope Science Institute, 3700 San Martin Dr. Baltimore,
MD
21218, USA\\
             \email{valenti@stsci.edu}
         \and
             Department of Astrophysical and Planetary Sciences, University of
Colorado, Boulder, CO 80309-0389, USA\\
             \email{dduncan@colorado.edu}
             }

   \date{Received 2003, Jun 25; accepted 2003, Oct 09}

   \abstract{
  We describe and analyze HST/STIS observations of the G2 V star
$\alpha$~Centauri~A (\object{$\alpha$~Cen~A}, \object{HD~128620}), a star similar to the
\object{Sun}. The
high resolution echelle spectra obtained with the E140H and E230H gratings
cover the complete spectral range 1133-3150 \AA\ with a resolution of 2.6
km\,s$^{-1}$,
an absolute flux calibration accurate to $\pm 5$\%, and an absolute wavelength
accuracy of 0.6--1.3 km\,s$^{-1}$. We present here a study of the E140H
spectrum
covering the 1140--1670~\AA\ spectral range, which includes   671 emission lines
representing 37 different ions and the molecules CO and H$_2$. For
\object{$\alpha$~Cen~A} and
the quiet and active \object{Sun}, we intercompare the redshifts, nonthermal line
widths, and parameters of two Gaussian representations of transition region
lines (e.g., \ion{Si}{iv}, \ion{C}{iv}), infer the electron density from the
\ion{O}{iv} intersystem lines, and compare their differential emission measure
distributions.
One purpose of this study is to compare the \object{$\alpha$~Cen~A} and solar UV spectra
to
determine how the atmosphere and heating processes in \object{$\alpha$~Cen~A} differ
from the
\object{Sun} as a result of the small differences in gravity, age, and chemical
composition of the two stars. A second purpose is to provide an excellent high
resolution UV spectrum of a solar-like star that can serve as a proxy for the
\object{Sun} observed as a point source when comparing other stars to the \object{Sun}. 
   }
\titlerunning{HST/STIS E140M spectrum of \object{$\alpha$~Cen~A}}
\authorrunning{Pagano et al.}

   \maketitle
%

 \section{Introduction}
Our knowledge and understanding of phenomena related to  magnetic activity in
late-type stars is based largely on the analysis of observations of  the \object{Sun}
obtained with high spatial, spectral and temporal resolution. In particular,
the different heating rates and emission measure distributions of stellar
chromospheres and transition regions can be understood by comparing stellar UV
spectra with corresponding solar spectra.  However, as strange as this may at
first appear, we lack a true ``reference spectrum'' for the \object{Sun} observed as a
star for such comparisons. In fact, the existing solar UV spectra  provided by
instruments on the {\em Solar Maximum Mission (SMM)} and the {\em Solar and
Heliospheric Observatory (SOHO)} typically have moderate to high spectral
resolution, but do not represent a full disk average, have uncertain wavelength
and absolute flux calibrations, and consist of a stitching together of many
small parts of the UV spectrum obtained at different times.
Table~\ref{solaratlases}  summarizes the instrumental
characteristics of these data sets. For example, the UV spectral atlas obtained
with the {\em High Resolution Telescope and Spectrograph (HRTS)} rocket
experiment \citep{brekke93b} and the recent UV spectral atlas obtained with the
{\em Solar Ultraviolet Measurements of Emitted Radiation (SUMER)} instrument on
the {\em Solar and Heliospheric Observatory (SOHO)} \citep{curdt} have high
spectral resolution, but do not provide the solar irradiance (the \object{Sun} viewed as
a point source) for direct comparison with stellar spectra. On the other hand,
spectra of the \object{Sun} as a point source obtained with the {\em Solar-Stellar
Irradiance Comparison Experiment (SOLSTICE)} instrument on the {\em Upper
Atmospheric Research Satellite (UARS)} \citep{rottman93}, the {\em EUV Grating
Spectrograph} \citep{woods90}, and the {\em Coronal Diagnostic Spectrometer
(CDS)} on {\em SOHO} \citep{brekke00} do not have sufficient spectral
resolution to resolve the line profiles. 

\begin{table*}
\begin{center}
\caption{Ultraviolet spectral atlases of the \object{Sun} and \object{$\alpha$~Cen~A}}
\label{solaratlases}
\begin{tabular}{lccccc}
\hline
\hline
Instrument & Spectral    & Spectral   & Solar    & Flux        & Reference \\
Used       & Range (\AA) & Resolution & Location & Calibration &           \\
\hline
HRTS       & 1190--1730  & 0.05\AA        & quiet \object{Sun}   & $\pm 30$\%   & (1) \\
           &             &                & active \object{Sun}  &              & (1) \\
UVSP/SMM   & 1150--3600  &$\sim 100,000$  & disk center &              & (2) \\
SOHO/SUMER & 465--1610   & 17,770--38,300 & disk center & $\pm$20\%    & (3) \\
           &             &                & sunspot, CH & $\pm$20\%    & (3) \\
SOHO/CDS   & 150--800    & 0.3--0.6\AA    & slit on disk& not given    & (4) \\
           & 307--632    & 0.3--0.6\AA    & \object{Sun}-as-a-star& 15--45\%    & (5) \\
SOLSTICE/UARS & 1190--4200 & 1--2 \AA     & \object{Sun}-as-a-star & $\pm$5\%   & (6) \\
rocket EGS & 300--1100   & 2\AA           & \object{Sun}-as-a-star & $\pm$15\%  & (7) \\
STIS E140H & 1140--1670  & 114,000        & \object{$\alpha$~Cen~A}      & $\pm5$\%    &
(8) \\
 \hline
\multicolumn{6}{l}{(1) \citet{brekke93b}, ~(2) \citet{uvsp}, \citet{woodgate80}, ~(3) \citet{curdt},}\\
\multicolumn{6}{l}{(4) http://solg2.bnsc.rl.ac.uk/atlas/atlas.shtml,  ~(5) \citet{brekke00}, ~(6) \citet{rottman93},}\\
\multicolumn{6}{l}{(7) EUV Grating Spectrograph, \citet{woods90},  ~(8) \citet{stismanual}, \citet{bohlin01}.}\\
\end{tabular}
\end{center}

\end{table*}

One way to obtain a close approximation to a high resolution spectrum of the
whole \object{Sun} observed as a point source with excellent S/N, absolute flux
calibration, and wavelength accuracy is to observe a bright star with very
similar properties to the \object{Sun}. We have done this with the {\em Space Telescope
Imaging Spectrograph (STIS)} instrument on HST  \citep{woodg}, obtaining a very
high S/N  and high resolution ($R=\lambda/\Delta\lambda\approx 114,000$)
spectrum of the star \object{$\alpha$~Cen~A}, a nearby (d=1.34 pc) twin of the \object{Sun} with
the same
spectral type (G2~V).  Although there are some small differences in effective
temperature and metal abundances between \object{$\alpha$~Cen~A} and the  \object{Sun} (see
below), this
{\em STIS} spectrum of \object{$\alpha$~Cen~A} can be considered the best available
``reference
spectrum'' for the \object{Sun} viewed as a star, because it is a full disk average, has
excellent wavelength and flux calibration \citep{bohlin01}, and covers the
entire 1130--3100 \AA\ UV range with high S/N and within a short period of
time. 

\object{$\alpha$~Cen~A}B (G2 V + K1 V) is the binary system located closest to the Earth
(d=1.34 pc).   It shows an eccentric orbit (e\,=\,0.519) with a period of almost 80 years \citep{Pour02}. Actually $\alpha$ Cen is a triple star system. The third member of the system, \object{$\alpha$ Cen C} or \object{Proxima Cen}, is a M5.5~Ve flare star (V\,=\,11.05)  about 12\,000 AU distant from \object{$\alpha$ Cen} and   only d=1.29 pc from the Sun \citep{perry}.  Thanks  to the high apparent brightness  (V\,=\,-0.01 and V\,=\,1.33 for the A and B component, respectively) and large parallax of the $\alpha$ Cen 
stars, their surface abundances, other stellar properties, and astrometric
parameters are among the best known of any star except the \object{Sun}.  \citet{GeD00},
\citet{Moreletal00}, and \citet{Pour02} have  reviewed recent determinations of
the physical characteristics of \object{$\alpha$~Cen~A}B. According to 
\citet{Moreletal00} and references therein, \object{$\alpha$~Cen~A} has nearly the same 
surface temperature of the \object{Sun} (T$_{eff}$=5790$\pm$30~K), slightly lower
gravity than the \object{Sun} ($\log g$=4.32$\pm$0.05, i.e.  0.76 g$_{\sun}$), and a mass of
 1.16$\pm$0.03 M$_{\sun}$  -  which is probably an upper limit, given different estimates 
 reported in the literature starting from 1.08  M$_{\sun}$ \citep{GeD00}. The same authors give 
 a metal overabundance of $\sim$0.2 dex with respect
to  the \object{Sun}, but similar Li and Be abundances to the \object{Sun}. In
Table~\ref{abundance}  we list the \object{$\alpha$~Cen~A} abundances
used in this paper, which were compiled from \citet{F-G-2001} and
\citet{Moreletal00}.
The age of \object{$\alpha$~Cen~A} is controversial:
\citet{Moreletal00}
derive an age in the range 2.7-4.1~Gyr depending on the adopted convection
model, while \citet{GeD00} estimate an age in the range 6.8-7.6 Gyr.  One could
argue that \object{$\alpha$~Cen~A} is younger than the \object{Sun} on the basis that it is
formed of
metal enriched  material, but the larger radius and lower gravity compared to
the \object{Sun} argue that the star is more evolved and somewhat older than the \object{Sun},
even considering its somewhat larger mass. 
 A closer analog to the Sun is \object{18~Sco} (V\,=\,5.50), but this star is too faint to get 
high S/N high resolution  UV spectra with STIS.

\begin{table}[h]
\caption{Abundances of \object{$\alpha$~Cen~A} in log units.}
\label{abundance}
\begin{tabular}{llc|llc}
\hline
\hline
Atom & Abund. & Ref. & Atom & Abund. & Ref.\\
\hline
  H & 12.00 &  1 &  S &  7.33 &  3 \\     
 He & 10.93 &  1 & Cl &  5.50 &  1 \\ 
 Li &  1.30 &  2 & Ar &  6.40 &  3 \\  
 Be &  1.40 &  3 &  K &  5.12 &  3 \\  
  B &  2.55 &  3 & Ca &  6.58 &  1 \\
  C &  8.72 &  1 & Sc &  3.42 &  1 \\
  N &  8.22 &  1 & Ti &  5.27 &  1 \\
  O &  9.04 &  1 &  V &  4.23 &  1 \\
  F &  4.56 &  3 & Cr &  5.92 &  1 \\
 Ne &  8.08 &  3 & Mn &  5.62 &  1 \\
 Na &  6.33 &  3 & Fe &  7.75 &  1 \\
 Mg &  7.58 &  3 & Co &  5.20 &  1 \\
 Al &  6.71 &  1 & Ni &  6.55 &  1 \\
 Si &  7.82 &  1 & Cu &  4.46 &  4 \\
  P &  5.45 &  3 & Zn &  4.85 &  4 \\
 \hline
\multicolumn{6}{l}{ References:}\\
\multicolumn{6}{l}{ 1) Feltzing \& Gonzalez(2001);}\\
\multicolumn{6}{l}{ 2) Morel et al.(2000);}\\
\multicolumn{6}{l}{ 3) Solar values from Grevesse \&
Sauval (1998);}\\
\multicolumn{6}{l}{ 4) scaled from the Fe abundance.}\\
\end{tabular}
\end{table}

\object{$\alpha$~Cen} has been extensively studied in the ultraviolet by IUE. \citet{Jordanetal} 
used IUE data to create simple one-dimensional models of the atmospheric structure of the two stars. 
\citet{Hallametal} have studied the rotational modulation of the most prominent lines in IUE spectra 
of \object{$\alpha$~Cen~A} and found a rotation period of about 29 d.  This is  consistent with the 
\citet{Boesgaard} estimate that the $\alpha$ Cen A rotation period is 10\% larger than the solar one, 
but is larger  than the  $\sim$22 d rotation period derived from  the 2.7$\pm$0.7 km\,s$^{-1}$
rotational velocity measured by \citet{saarosten}, assuming a radius of
$\sim$1.2
R$_{\sun}$ and an orbital inclination of $\sim$79$\degr$.
\citet{Ayresetal}   have studied the time variability of the most prominent UV lines of \object{$\alpha$~Cen~A} 
and B during about 11 years of observations.  While a clear evidence of a solar-like activity cycle 
was found for \object{$\alpha$~Cen~B}, UV line fluxes  from \object{$\alpha$~Cen~A} do not give any clear 
indication for an activity cycle.  

In this paper we report on the \object{$\alpha$~Cen~A} spectrum recorded with the E140
grating
by {\em HST/STIS} between 1140--1670 \AA, while the analysis of the E230H
spectrum (1620--3150 \AA) will be published in a forthcoming paper.  Information on data acquisition and 
reduction are
provided in Section~\ref{data}, the  spectral line identification and  the
analysis of interesting lines are presented in Section~\ref{results}. A
detailed comparison of our {\em STIS} \object{$\alpha$~Cen~A} spectrum, with the {\em
SOHO/SUMER} \citep{curdt} and the {\em SMM/UVSP} \citep{uvsp} spectra of the
\object{Sun} is given in Section~\ref{sun-comp}. Then, we derive the \object{$\alpha$ Cen A} transition 
region electronic densities (Section~\ref{density}),  and emission measure distribution 
(Section~\ref{emissionmeasure}). In Section~\ref{ism-sec} we call the reader's attention on some 
absorption features present in high exicitation lines, and  give our conclusions in Section~\ref{conclu}.

 \section{The \object{$\alpha$ Cen A} Data}
 \label{data}

The E140H spectrum of \object{$\alpha$~Cen~A} was acquired on 1999 Feb 12 with 3
exposures of 4695~s each, centered at 1234, 1416, and 1598~\AA, respectively.
The E140H mode ensures an average dispersion of $\lambda /228,000$~\AA\ per
pixel, which corresponds to a resolving power of 2.6 km\,s$^{-1}$. The E140H
grating
is  used with the FUV-MAMA detector, which we operated in TIME-TAG mode. We
used
the 0.2$\times$0.09 arcsec aperture.

The data were reduced using the {\em STIS} Science Team's IDL-based software,
CALSTIS (Version 6.6). CALSTIS performs a variety of functions including flat
fielding, assignment of statistical errors, compensation for the Doppler shifts
induced by the spacecraft's motion in orbit, conversion of counts to count
rates, dark-rate image subtraction, and the removal of data from bad/hot
pixels.  Wavelength calibration was carried out assuming the post launch
echelle  dispersion coefficients and a dispersion coefficient correction for
the Monthly  MSM offsets released to the {\em STIS} Science Team on 1999
September \citep{lind99b}. The on-board Pt lamp spectra taken in association
with the science observations were used to measure zero point adjustments. For
the echelle observations, CALSTIS computes a wavelength offset for each
spectral order.  The adopted offset is the median of  these offsets.  As a
check for the success of the algorithm used, we have  verified that the
measured offsets are all within one pixel of the median offset.  As a further
check on the accuracy of the wavelength scale, we measured the centroids of
emission lines recorded in adjacent orders, and found that the results agree to
within less than 1 pixel. The nominal  absolute wavelength  accuracy is 0.5--1
pixel (i.e. 0.6--1.3 km\,s$^{-1}$) \citep{stismanual}. 

CALSTIS outputs a file containing wavelength, flux, and error vectors, which is
 used in all subsequent processing.  To remove the effects of scattered light
that are important near the Ly-$\alpha$ line, we used the IDL ECHELLE\_SCAT
routine \citep{lind99a} in the {\em STIS} Science Team's software package.
This routine uses  the first estimate of the spectrum and a scattering model of
the spectrograph to  determine the intensity of the scattered light and to
estimate what the spectrum  plus scattered light image should look like.
Comparison of this calculated  spectrum with the observations yields
differences that indicate the errors in  the first estimate of the spectrum.
This spectrum is then corrected and the  process is iterated until acceptable
agreement is obtained between the  prediction and the observed image.

After correction for scattered light, the spectrum was then analyzed using
software packages  written in IDL.  We used routines of the ICUR fitting
code\footnote{ICUR (http://sbast3.ess.sunysb.edu/fwalter/ICUR/icur.html) is a
general purpose screen-oriented data analysis program written in IDL  for
manipulating and analyzing one dimensional spectra. It is  distributed to the
public by the authors  (F.M. Walter and J.E.  Neff).}, adapted to handle our
{\em STIS} data, which perform  multi-Gaussian fits to the line profiles using
\citet{bw} CURFIT algorithm.  To correct for instrumental broadening, we
convolved each proposed fit to an emission line profile  with the instrumental
line spread function (LSF), which was assumed to be a  Gaussian with the
nominal width ranging from $\sim$1.2 pixel at 1200 \AA\  to  $\sim$1 pixel at
1700 \AA\ \citep{stismanual}, as is appropriate for lines which  are  much
broader than the width of the LSF.

 \section{Results}
 \label{results}

 \subsection{The Ultraviolet Spectrum of \object{$\alpha$ Cen A}}
 \label{spectrum}

\begin{figure*}[ht]
\begin{centering}
\resizebox{15cm}{20cm}{\includegraphics{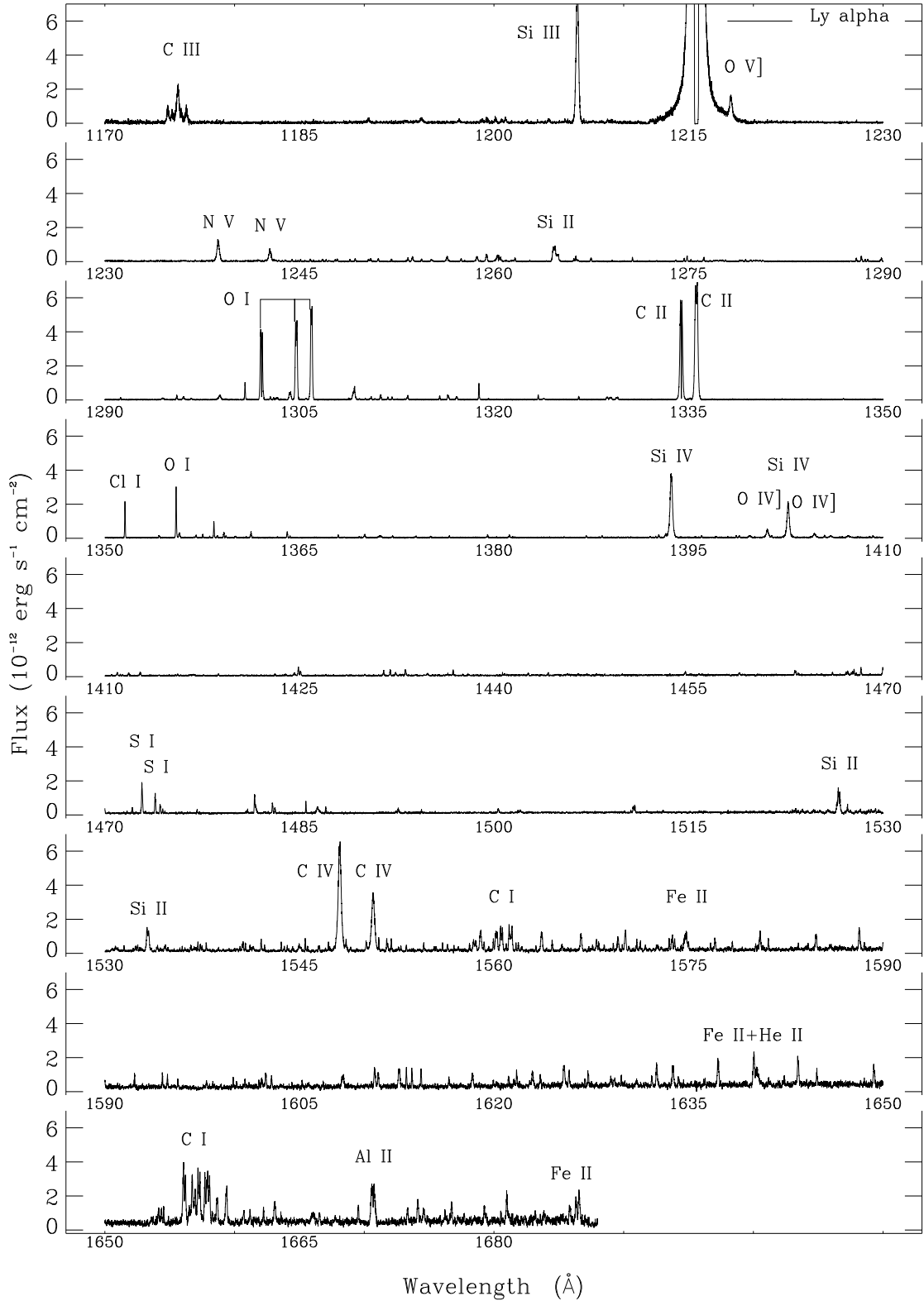}} 
\vspace*{1cm}
 \caption{The E140H spectrum of \object{$\alpha$ Cen A} obtained on 1999 Feb 12. 
Important spectral features are marked.\label{fig1}}
 \end{centering}
 \end{figure*}

In Figure~\ref{fig1} we show  the E140H spectrum of \object{$\alpha$~Cen~A}. We have
measured a
total of 662 emission features of which 77 are due to blends of  two or more
lines, 71 are due to unidentified transitions, and 514  are identified as due
to single emission lines.  Taking into account  the  157 lines identified in
blended features, we find a total of 671 emission lines in this spectrum. In
Table~\ref{stis} we list all the ions that have been identified. Most of these
lines are due to  \ion{Si}{i}, \ion{Fe}{ii}, \ion{C}{i}, which together
contribute 441 lines, but \ion{S}{i} and \ion{Ni}{ii} are each represented by more than 30 lines.

 \begin{table}[h]
 \caption{ Lines detected in  the STIS E140H spectrum of \object{$\alpha$ Cen A}.}
 \label{stis}
 \begin{tabular}{lrr|lrr}
 \hline
\hline
 Ion &\multicolumn{2}{c|}{No. of lines}&Ion &\multicolumn{2}{c}{No. of lines}
 \\
&       Total   &Blended        &               &       Total   &Blended \\
 \hline
 Si I    &155   &46     &O IV   &5  &1  \\
 Fe II  &144    &20     &Ca II  &4  &3  \\
 C I    &142    &32     &Cl I   &3  &   \\
 S I    &55     &6      &S II   &3  &\\
 Ni II   &34    &8      &S III  &3  &1  \\
 Si II  &17     &4      &S IV   &3  &1  \\
 H$_2$  &14     &11     &C IV   &2  &  \\
 N I     &13    &6      &Fe V   &2  &   \\
 CO     &11     &6      &N V    &2  &   \\
 C III  &8      &1      &O III  &2  &   \\
 Fe IV  &8      &1      &O V    &2  &   \\
 Si III &7      &2      &Si IV  &2  &   \\
 O I    &6      &1      &S   V  &2  & 1  \\
 C II    &6      &4      &      &  &   \\
 P I    &5      &       &       & &  \\
 \hline
 \multicolumn{6}{l}{1 line each for Al II, Al IV, Ca VII, Cl II, Cr II,}\\
 \multicolumn{6}{l}{~~~~~~~  Fe XII, H I, He II, and N IV}      \\

 \multicolumn{6}{l}{1 line in blended features for P II, and Si VIII}   \\
 \hline
 \end{tabular}
 \end{table}

Table~4\footnote{Table~4 is also available at the CDS},
lists the line identifications, laboratory and measured wavelengths,
radial velocity shifts corrected for the  stellar radial velocity of
--23.45 km\,s$^{-1}$, computed using the orbital parameters and
ephemeris given  by \citet{Pour02}, line full-widths at half-maxima
(FWHM), and line fluxes.  The laboratory wavelengths listed in
Table~4 are from \citet{san86}, unless otherwise noted in the
table.  We used Gaussian fits to the line profiles to measure
wavelengths, FWHM, and fluxes for single or blended emission lines
which do not show central reversals. For the lines which have
interstellar absorption components or central reversals, i.e.  the most
intense optically thick chromospheric lines of \ion{C}{i}, \ion{O}{i},
\ion{Si}{ii}, and \ion{C}{ii}, we instead integrated the flux contained
in a suitable wavelength interval and tabulated the FWHM of the
observed profile.  In Table~4 these lines are indicated with
``CR'' in the {\em Notes} column.

The strongest transition region lines show broad wings, and therefore
do not have a Gaussian profile. For these lines, we list in
Table~4 the line centroid, the FWHM of the observed profile
and  the flux integrated in a suitable wavelength interval.  The
analysis  of these lines is reported in Section~\ref{BCNC}.

Absorption features  due to the interstellar medium have been
measured  in a number of lines originating in   transitions from the
ground level.   Such lines  are indicated  with ``ISM'' in
Table~4 (column {\em Notes}).  They will be  discussed in a
separate paper, together with the derived properties of the
interstellar medium along this line of sight.

Several intersystem lines  are present in the spectrum of the
\object{$\alpha$~Cen~A}, including the  \ion{O}{iv} UV 0.01
intercombination  multiplet $2s^22p^2P^0_J-2s2p^{2~4}P_J$, that are
diagnostics  of electron density (cf.  \citealt{dlm02, bjb96}, and
references therein), the \ion{N}{iv} line at 1486~\AA, and the
\ion{O}{iii} line at 1666~\AA. We have used these lines to measure
densities in the \object{$\alpha$~Cen~A} chromosphere and  transition
region as discussed in Section~\ref{density}.

\subsection{Comments on individual line identifications}
\label{comm}
\begin{figure*}[ht]
\begin{centering}
\resizebox{15cm}{13cm}{\includegraphics{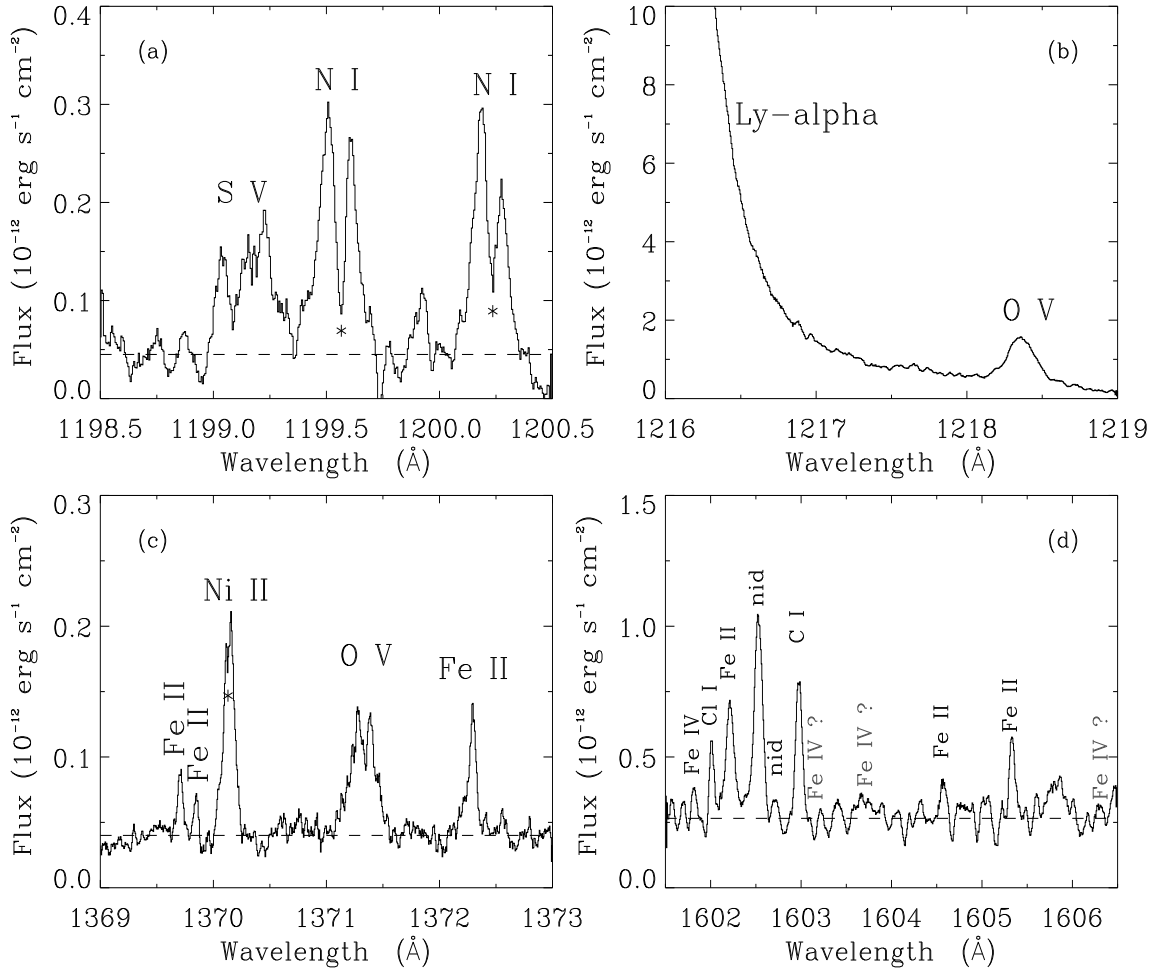}} 
\caption{Blow-up of the regions  containing the \ion{S}{v} 
1199.134 \AA\  line ({\em panel a}), 
the \ion{O}{v} 1218 \& 1371 \AA\ lines ({\em panels b and c}), and the 
\ion{Fe}{iv}  
quadruplet between 1601.5 and 1606.5 \AA\ ({\em panel d}).  Of this   
quadruplet we could measure only the 1602 \AA\  line. Light-ink labels in 
panel {\em  d} indicate the positions of the missed \ion{Fe}{iv}  lines. 
The  symbol $*$ in  panels {\em a} and {\em c} marks the absorption components 
due 
to the interstellar medium. In all of  
the panels the wavelength scale has been 
shifted according to the stellar radial velocity. 
\label{nuovo}}
\end{centering}
\end{figure*}

According to the NIST database, we have identified the broad feature near
1199~\AA\ as \ion{S}{v} (see Figure~\ref{nuovo}a). 
However, it is possible that other unidentified lines
are present. In fact, the  flux measured at 1199.08 \AA\  seems too large to be consistent with
the differential emission measure distribution derived in
Section~\ref{emissionmeasure}. 

 The chromospheric Ly$\alpha$ emission line is altered greatly by the
superimposed narrow, weak deuterium (D I) interstellar absorption and by
very broad, saturated hydrogen (H I) interstellar, heliospheric, and
astrospheric absorption, and by geocoronal emission.
The Ly$\alpha$ line flux given in Table~4   was estimated by fitting a Gaussian to the 
wings of the line profile, disregarding the central part of the line, which is strongly affected 
by ISM absorption and geocoronal emission, and including a second Gaussian to account for 
the Deuterium absorption. This is a very rough estimate of the Ly$\alpha$ flux.  We refer to 
the \citet{Linsky96} and \citet{w01} papers for reliable estimates of the intrinsic Ly$\alpha$ 
in $\alpha$~Cen~A.

The two \ion{O}{v} lines that we have measured in the \object{$\alpha$~Cen~A} {\em STIS}
spectrum have radial velocities differing by about 1.8~km\,s$^{-1}$, with the
1218~\AA\
line less red-shifted than the line at 1371~\AA. On the \object{Sun}, the 1371~\AA\ line
has a Doppler shift of $\sim5$ km\,s$^{-1}$ greater than the 1218~\AA\ line,
but
\citet{Brekke93} concluded that such a difference between the two lines can be
explained only by an error in the adopted laboratory wavelength of the
\ion{O}{v} 1218~\AA\ line, which is an intersystem line and thus difficult to
measure in the laboratory. However, if this were the case, adoption of the
wavelength 1218.325~\AA\ suggested by \citet{Brekke93}  as the laboratory
wavelength, leads to a significant difference (2.8 km\,s$^{-1}$) in the
opposite sense.
We suggest that the main reason of the slight wavelength disagreement, even on
the \object{Sun}, can be attributed to the difficulty in measuring the wavelength of the
\ion{O}{v} 1218~\AA\ line (see Figure~\ref{nuovo}b) 
in the sloping wing of the Ly$\alpha$ line. The
\ion{O}{v} line at 1371~\AA\  (see Figure~\ref{nuovo}c)  
shows a double peak with an apparent central
reversal. We know of no explanation for this effect  as the line is unlikely 
to be optically thick and thus self-reversed, and interstellar absorption is
also unlikely.

\begin{figure}[h]
\begin{centering}
\resizebox{9cm}{7cm}{\includegraphics{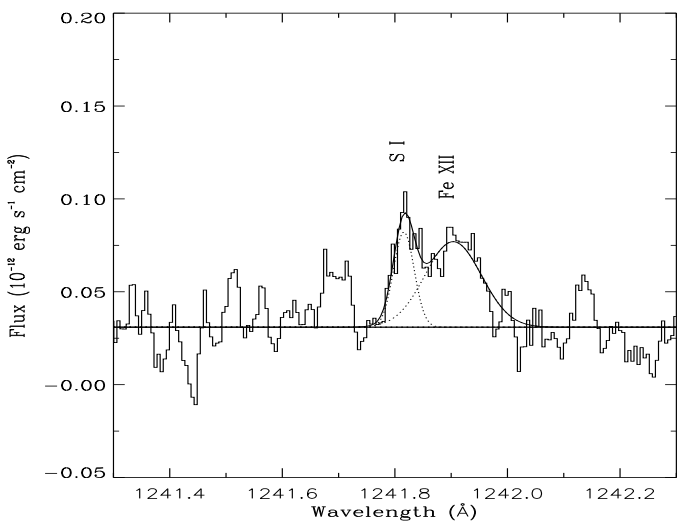}}
\caption{A double-peak structure  identified as  the \ion{S}{i} 
at 1241.9 \AA\ and the \ion{Fe}{xii} 1242~\AA\ lines.}
\label{fe12}
\end{centering}
\end{figure}

A blow-up of the region with  a complex feature  located near  
1241.8 \AA\ is
shown in Figure~\ref{fe12}. The feature is noisy, but its double-peak structure 
is preserved even  after  smoothing with a boxcar average
of width as  large  as 13 pixels.  We have therefore fitted the profile
with two Gaussians, and identified the two lines as 
the  \ion{S}{i}   1241.9 \AA\ and   \ion{Fe}{xii} 1242~\AA\ lines.
Since the   \ion{Fe}{xii} line is formed at temperature $\log T=6.13$, its
 predicted  
thermal width  is $\sim$33 km\,s$^{-1}$. We have frozen the line width of the
\ion{Fe}{xii} line to its thermal width, and derived  a flux of 
6.3$\times$10$^{-15}$ erg s$^{-1}$ cm$^{-2}$. An a-posteriori check for the 
 accuracy of our measured flux 
is given by the excellent agreement between the 
emission measure derived by using this line at $\log T=6.13$ and the emission 
measure derived
at temperatures $\log T=6.04$ and 6.3 from Chandra spectra \citep{raassen} (cf. 
Section~\ref{emissionmeasure}).

The weak emission feature observed  in solar spectra 
at $\sim$ 1356.88 \AA\ was tentatively
attributed to the \ion{S}{iii} line at 1357.0 \AA\ by \citet{feld75}. Its
laboratory wavelength makes this line slightly blue-shifted, in contrast with
the expectation (cf. Section~\ref{vshift}), therefore we can argue that either
the  identification is wrong or the laboratory wavelength given by
\citet{feld75} is inaccurate.

\begin{figure*}
\begin{centering}
\resizebox{8.0cm}{5cm}{\includegraphics[angle=90]{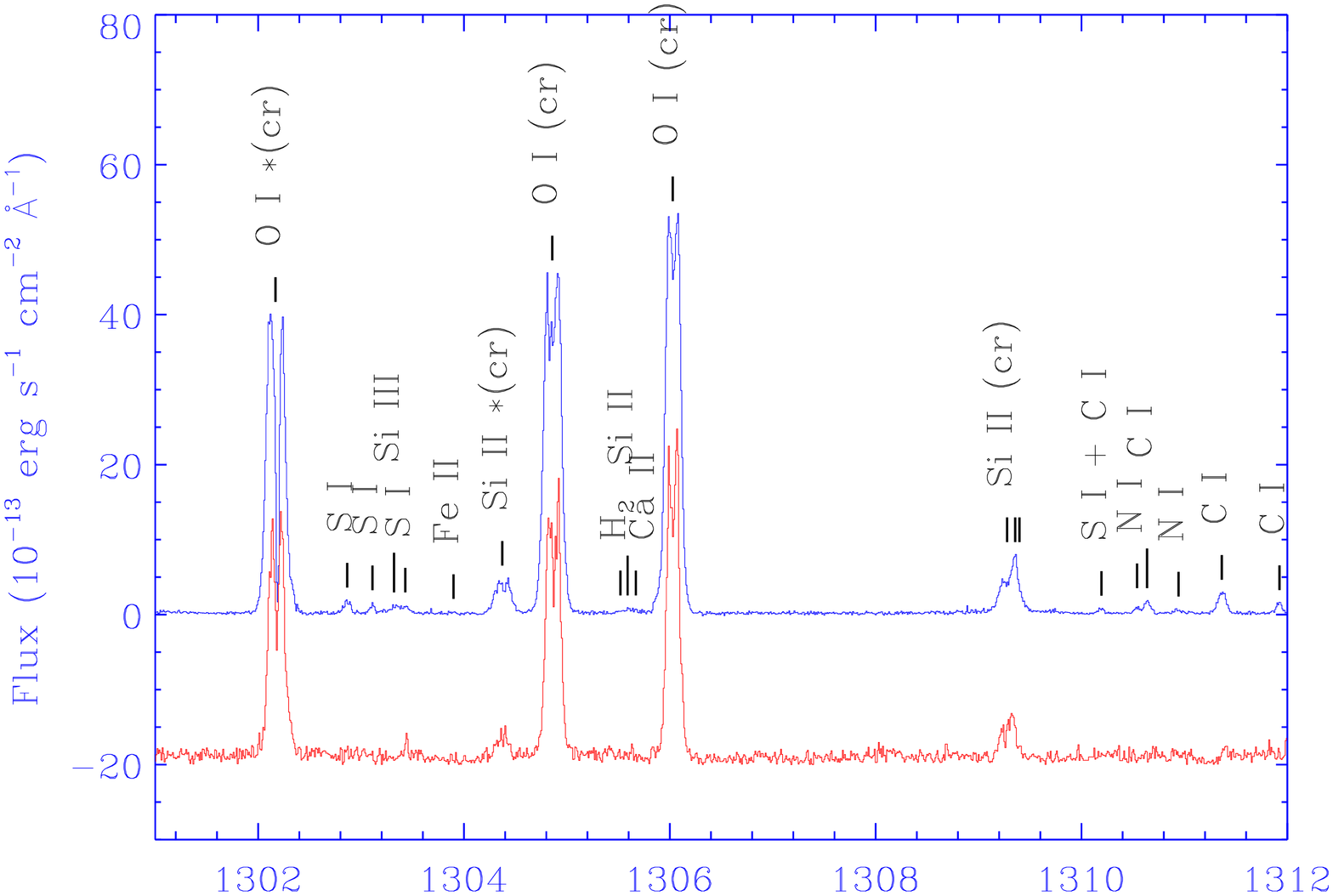}}
\resizebox{8.0cm}{5cm}{\includegraphics[angle=90]{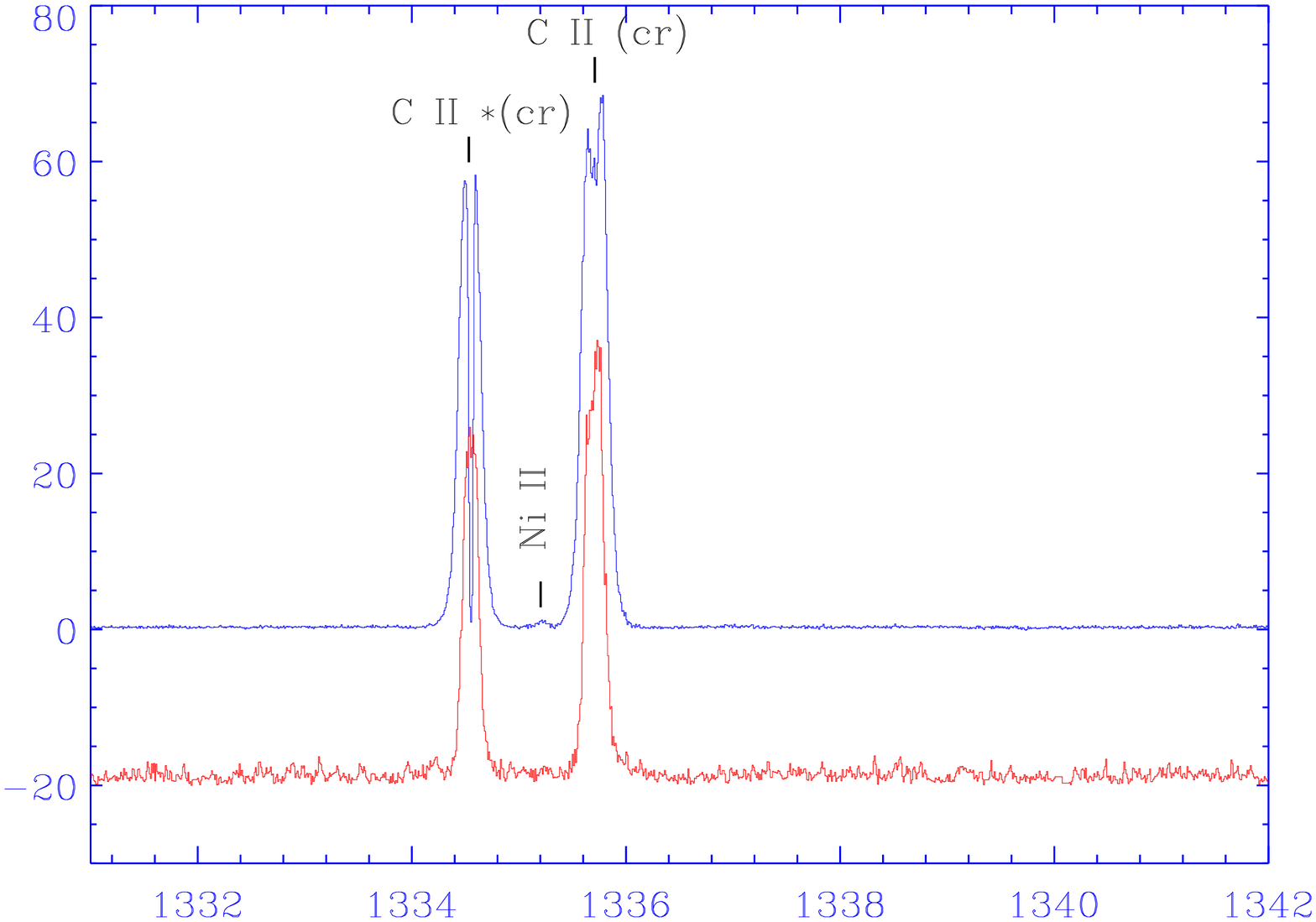}}\\
\resizebox{8.0cm}{5cm}{\includegraphics[angle=90]{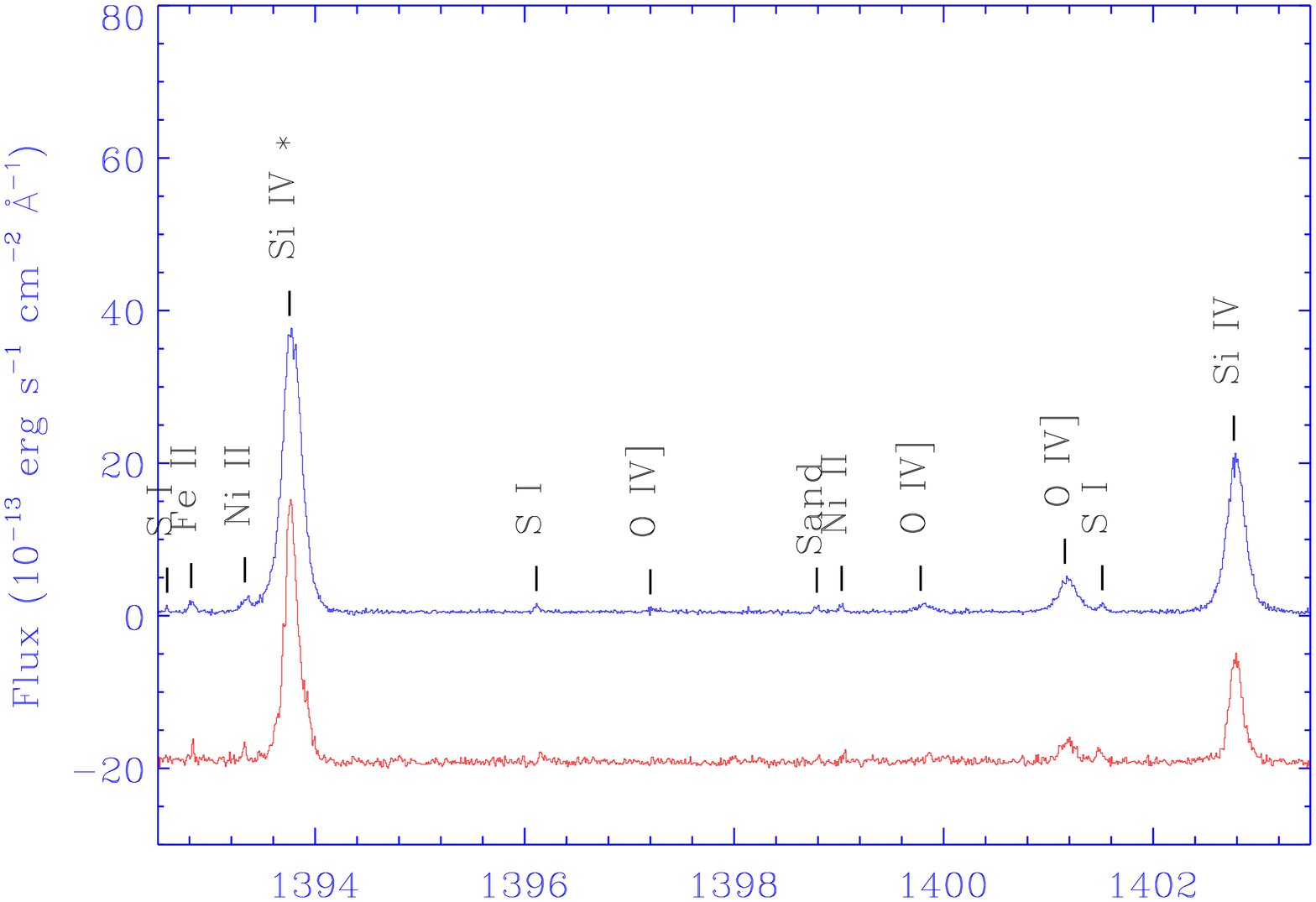}}
\resizebox{8.0cm}{5cm}{\includegraphics[angle=90]{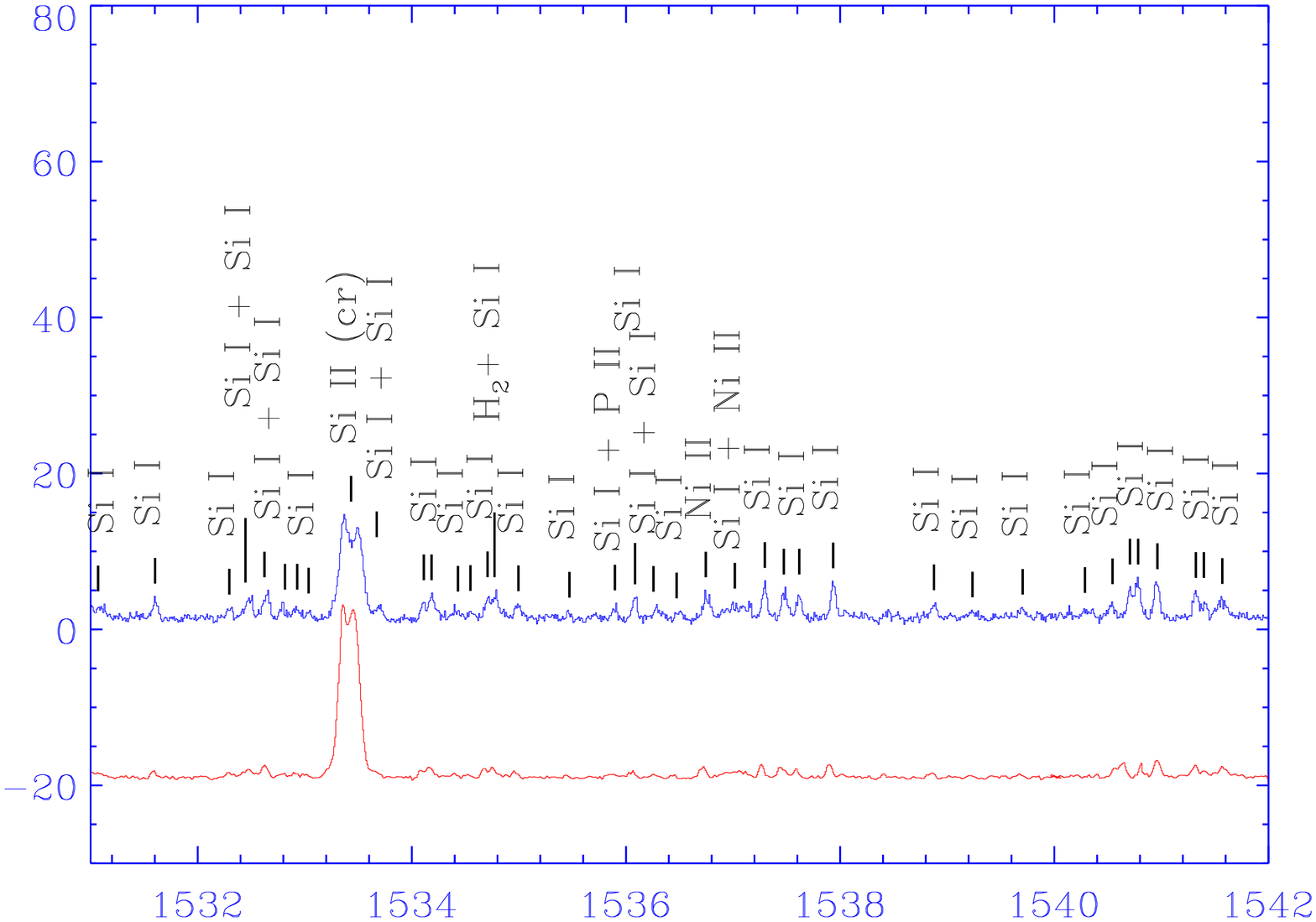}}\\
\resizebox{8.0cm}{5cm}{\includegraphics[angle=90]{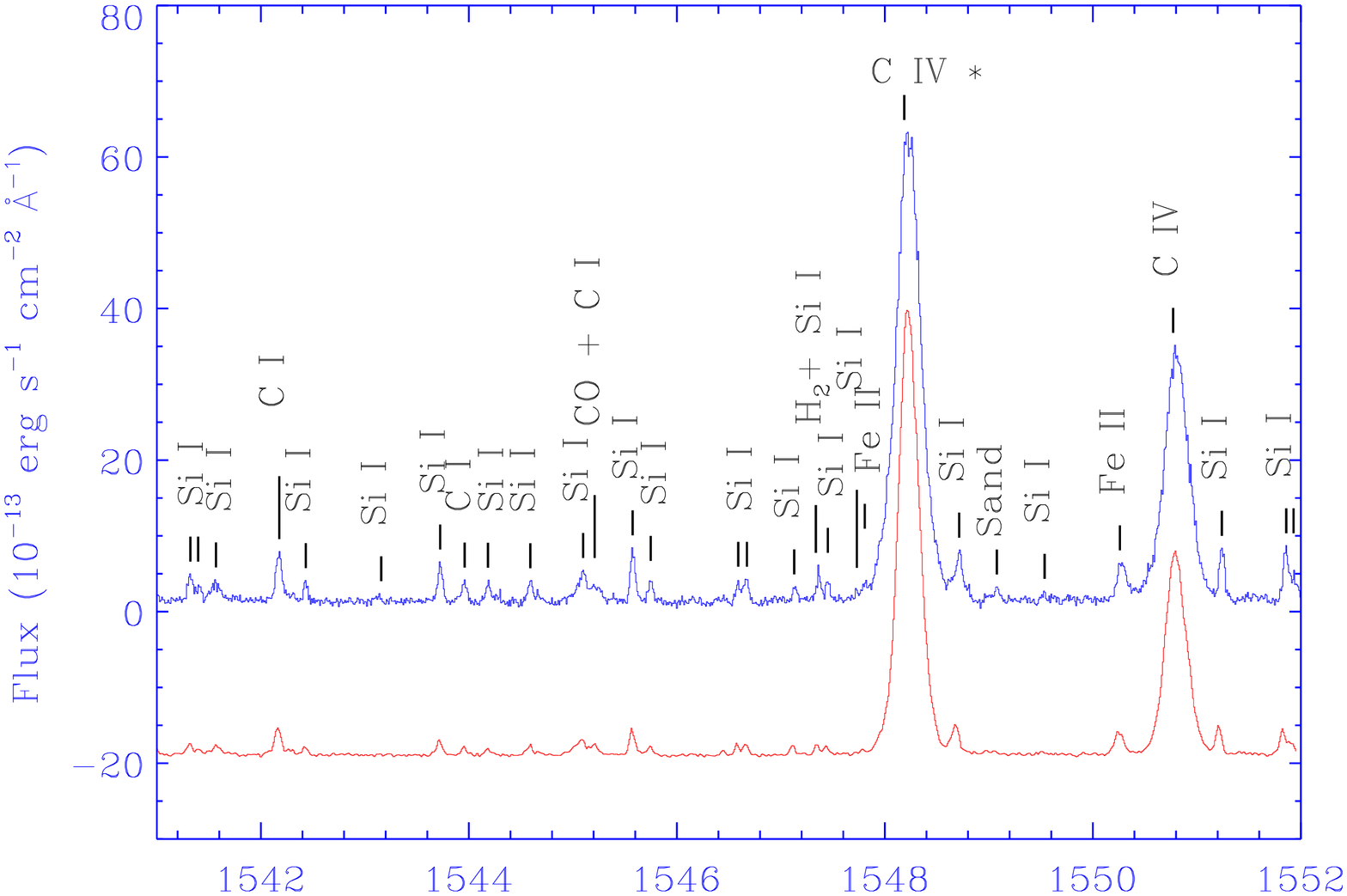}}
\resizebox{8.0cm}{5cm}{\includegraphics[angle=90]{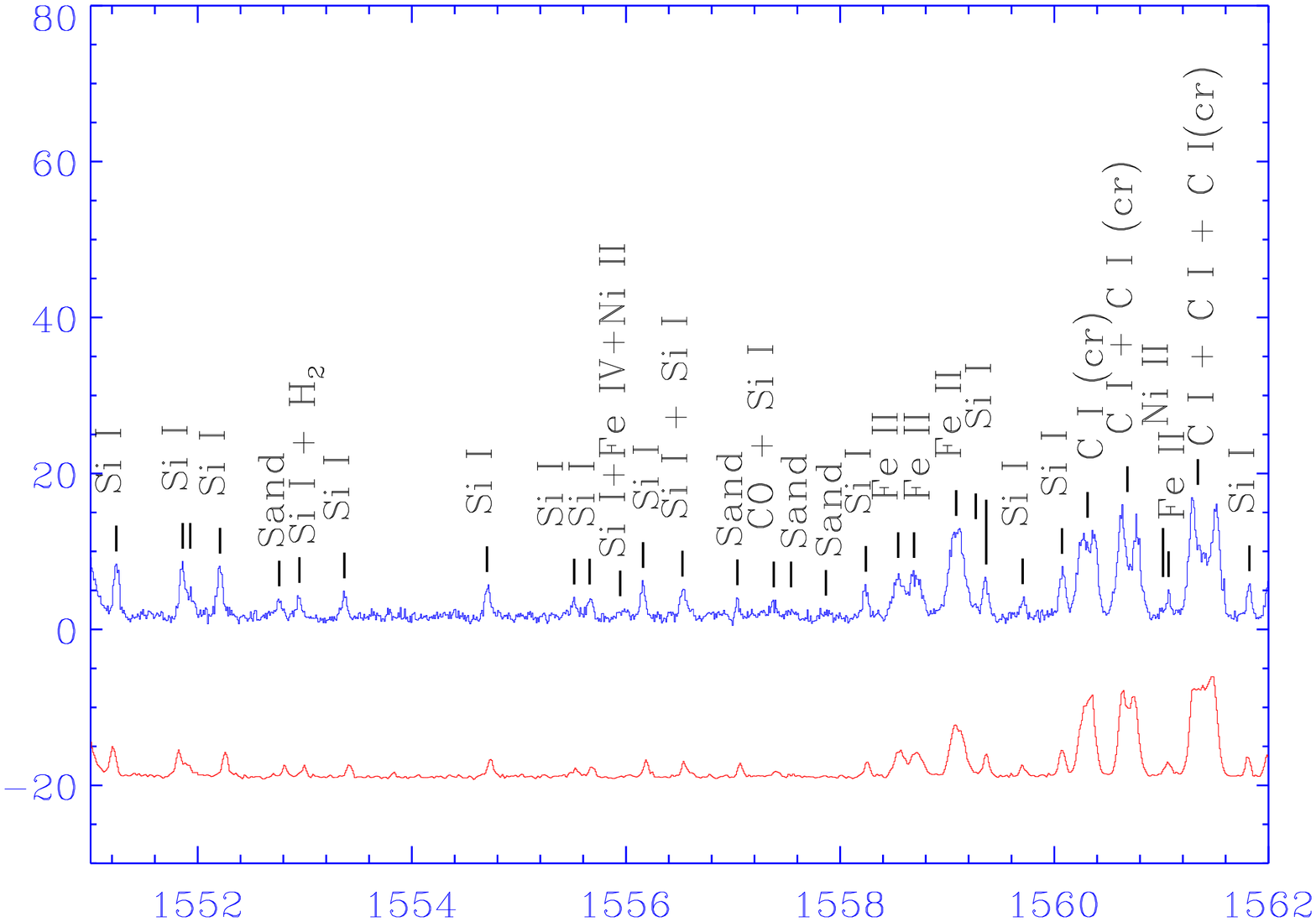}}\\
\resizebox{8.0cm}{5cm}{\includegraphics[angle=90]{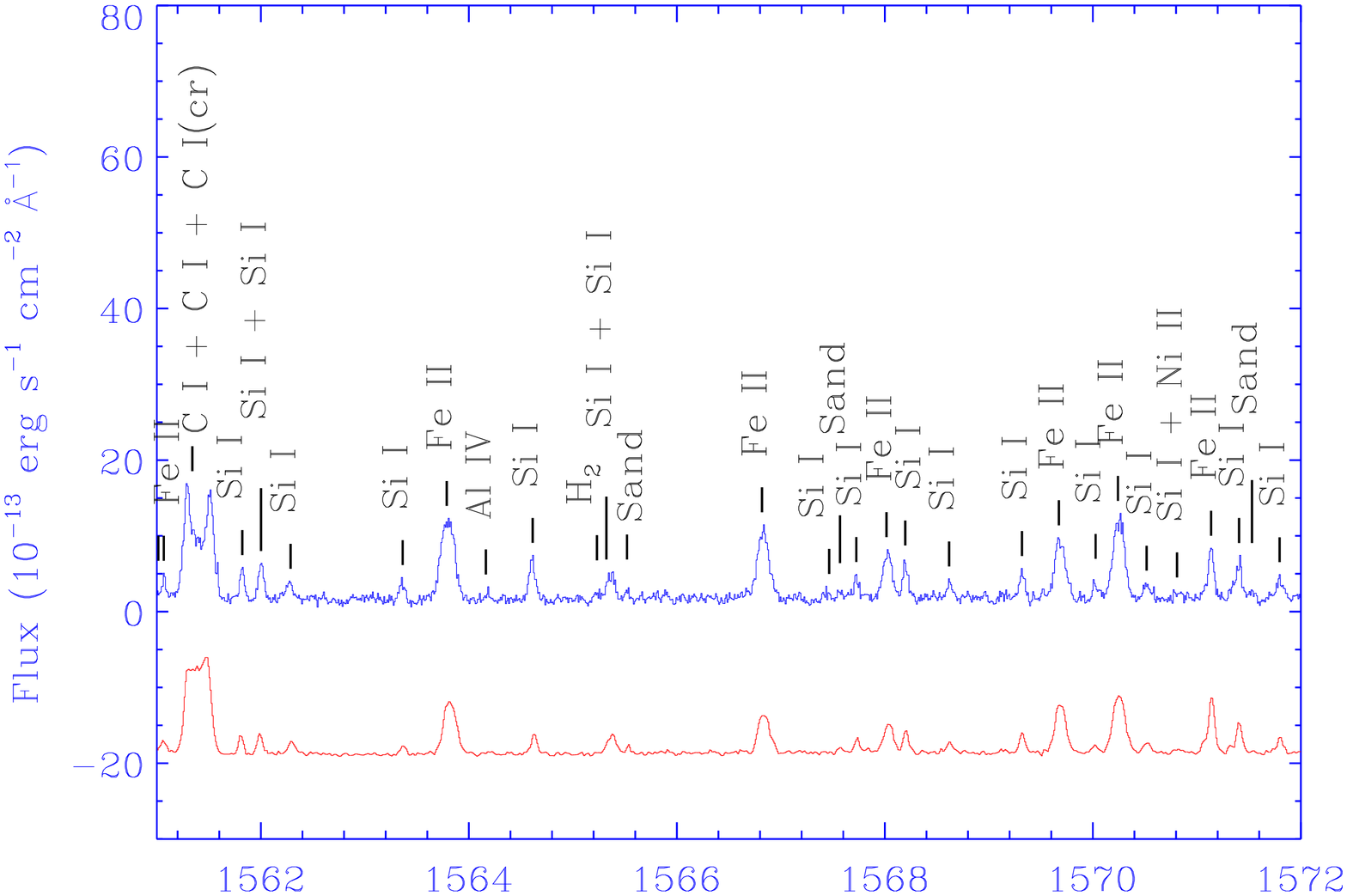}}
\resizebox{8.0cm}{5cm}{\includegraphics[angle=90]{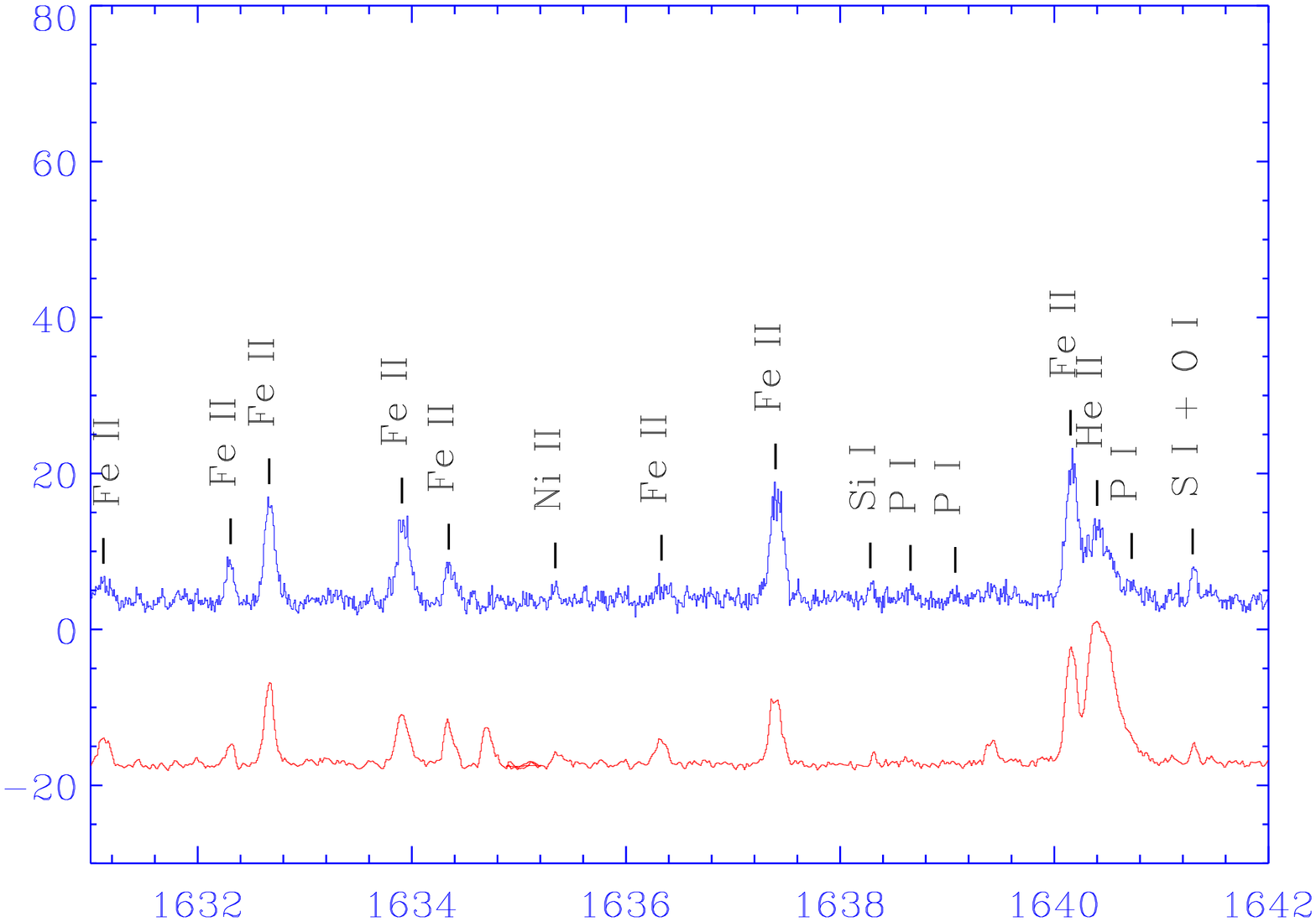}}\\
\resizebox{8.0cm}{5cm}{\includegraphics[angle=90]{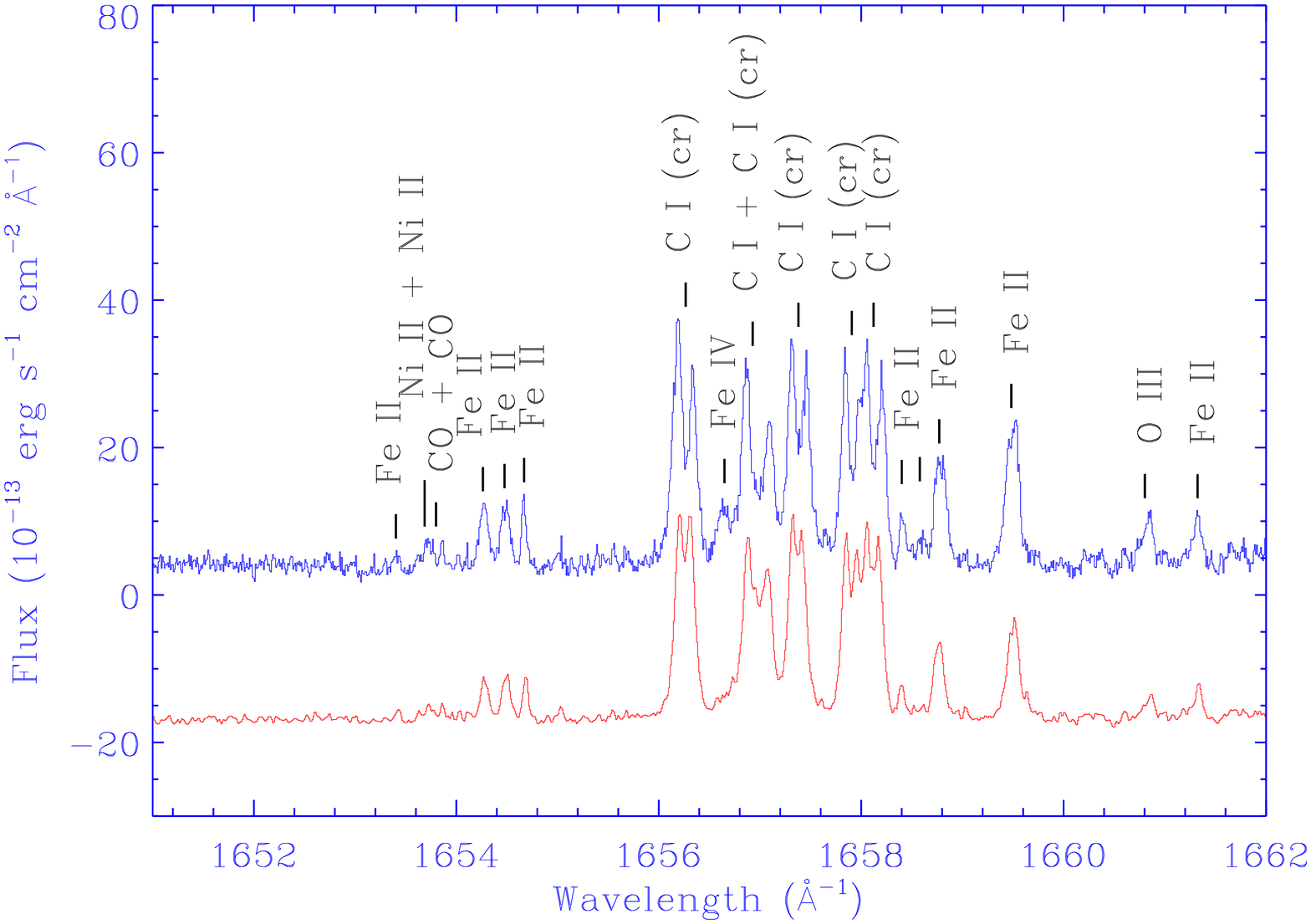}}
\resizebox{8.0cm}{5cm}{\includegraphics[angle=90]{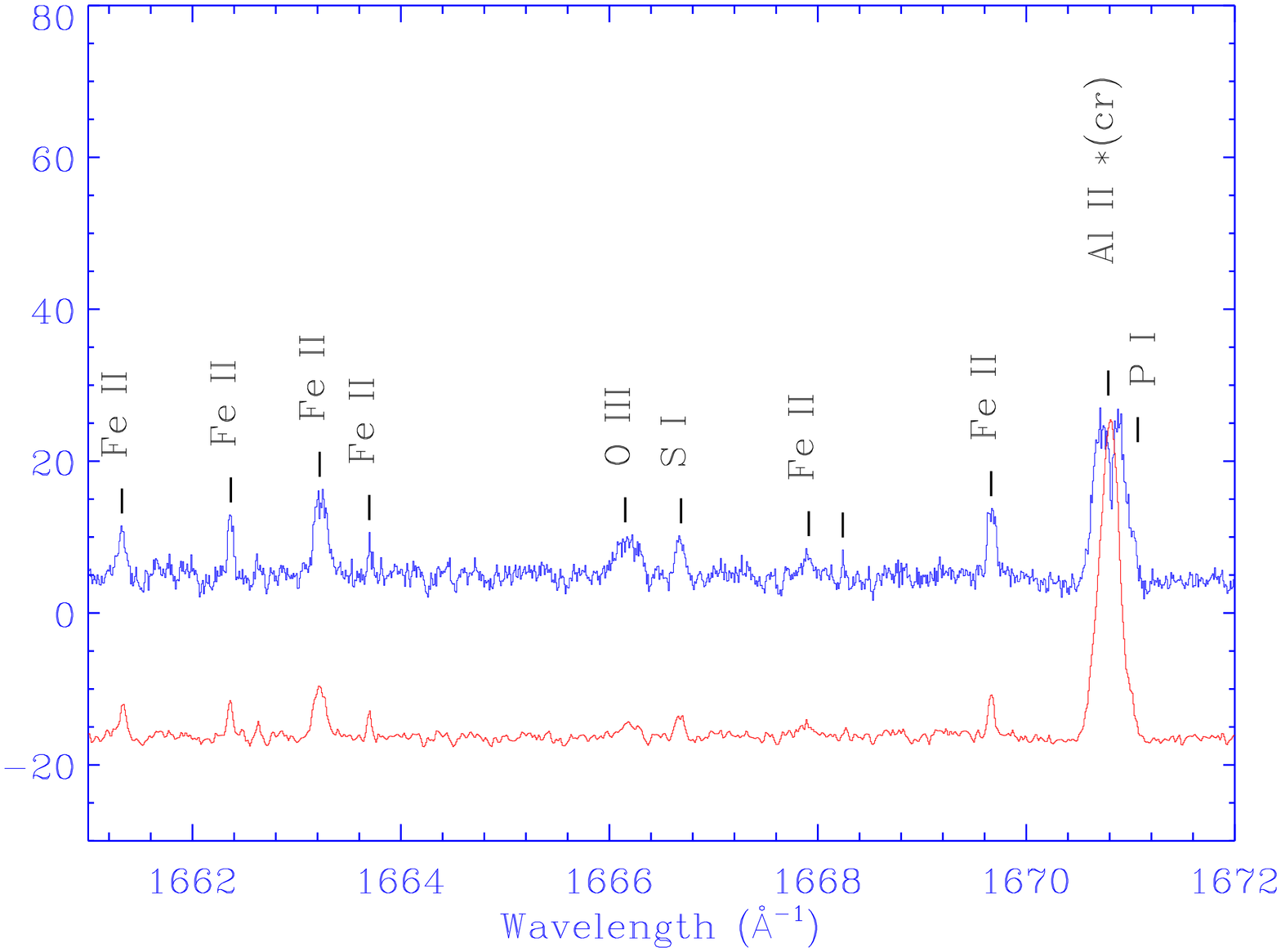}}\\
\caption{Plots of interesting portions of the \object{$\alpha$ Cen A}  HST/STIS and solar SMM/UVSP spectra, the latter shifted of $-$2e-12 erg s$^{-1}$ cm$^{-2}$ \AA$^{-1}$. Important spectral features are marked.}
\label{uvsp-stis}
\end{centering}
\end{figure*}

We have measured the \ion{Fe}{iv} line at 1602 \AA\ that belongs to a multiplet
of four lines. A careful inspection of the spectrum shows  slight flux
increments at the  wavelengths corresponding to the 1603.181, and 1603.730~\AA\
\ion{Fe}{iv} lines, which, however, are below our detection limit as shown in 
Figure~\ref{nuovo}  (panel d),  but we do not
find any appreciable emission feature corresponding to the  fourth line of this
multiplet at 1606.333~\AA. While all of the \ion{Fe}{iv} lines have been
identified in the \citet{kelly} line database, no \ion{Fe}{iv} lines have been
identified in the solar spectrum analyzed by \citet{san86}. We have inspected
the solar {\em SMM/UVSP} spectrum \citep{uvsp} to look for \ion{Fe}{iv} lines,
but even the strongest line measured in the  \object{$\alpha$~Cen~A} {\em STIS} spectrum
at
$\sim$1656 \AA\ is missing in the solar spectrum, as shown in the left-bottom 
panel 
of Figure~\ref{uvsp-stis}. 

 \subsection{The Broad Wings of the Transition Region Emission Lines}
 \label{BCNC}

 \normalsize

\begin{figure}[h]
\begin{centering}
\resizebox{9cm}{11cm}{\includegraphics{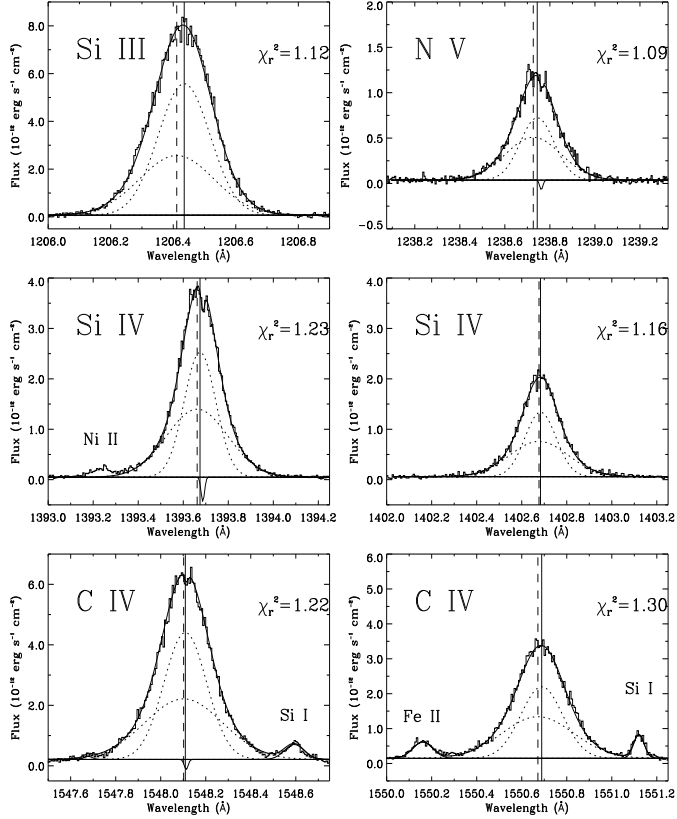}}
 \caption{The  \ion{Si}{iii},  \ion{N}{v}, \ion{Si}{iv}, and \ion{C}{iv} line
profiles. The narrow and broad dashed lines indicate the narrow and broad Gaussian 
components, respectively, required to best fit the
broad wings of these transition region lines. The vertical solid and dashed
lines indicate the centroids of the narrow and broad Gaussians, respectively.} 
 \label{fig3}
\end{centering}
 \end{figure}

As shown by \citet{w97}, the strongest transition region emission lines of
\object{$\alpha$~Cen~A} have  profiles  
with broad wings. We find that broad wings are present in
the \ion{Si}{iii} $\lambda$\,1206~\AA, \ion{N}{v} $\lambda$\,1238~\AA,
\ion{Si}{iv} $\lambda$\,1393 \& 1402~\AA, and \ion{C}{iv} $\lambda$\,1548 \&
1502~\AA\ line profiles. For these lines we used one narrow Gaussian component
(NC) to fit the line core and one  broad Gaussian component (BC) to fit the
broad wings (see  Figure~\ref{fig3}). This bi-modal structure of the transition
region  lines is typically observed for several RS CVn-type stars (i.e.,
\object{Capella} and  \object{HR~1099}), main sequence type stars (i.e., \object{AU Mic}, \object{Procyon},
\object{$\alpha$~Cen~A}, and \object{$\alpha$~Cen~B}), and the  giants \object{31~Com}, \object{$\beta$~Cet},
\object{$\beta$~Dra}, \object{$\beta$~Gem}, and \object{AB~Dor} \citep{lw94,linsky95,pagano00}.
\citet{w97} showed that the narrow  components can be produced by turbulent
wave dissipation or Alfv\'en wave heating mechanisms, while the broad
components, that resemble the explosive  events on the \object{Sun}, are diagnostics for
microflare heating. Analysis of {\em SUMER} data led \citet{peter01} to propose
an alternative explanation for the broad Gaussians, which he calls the ``tail
component'', seen in lines formed at temperatures between 50,000 and 300,000~K
in the chromospheric network. He argues that the tail component originates in
coronal funnels that magnetically connect the lower transition region with the
corona, and the broadening is by passing magneto-acoustic waves. 

Table~\ref{TtwoG} lists the parameters resulting from our multi-Gaussian
fits\footnote{For the fits of \ion{N}{v} 1238 \AA, \ion{Si}{iv} 1393 \AA, and
\ion{C}{iv} 1548 \AA, a third Gaussian component was used to account for the
absorption feature   possibly 
originating in the intervening interstellar medium.}.  Both
the narrow and broad components are redshifted with respect to the stellar
chromosphere, whose rest velocity is determined by the mean velocity of 80
selected Si I lines as discussed in Section \ref{vshift}. The narrow components
show larger redshifts as is seen in solar data \citep{peter01}. This effect
was also noticed by \citet{w97}, who analyzed the Si IV 1393~\AA\ line in
GHRS/HST spectra of \object{$\alpha$~Cen~A}. 

The broad and narrow Gaussian components have comparable intensity as
the flux-weighted mean ratio between the flux in the broad component and the
total flux is $0.46\pm 0.05$. This ratio is typical for the  most active
stars studied by \citet{w97}, and it appears to be independent of the 
activity level of the star.

\setcounter{table}{4}
\begin{table}
 \caption{Parameters derived from the multi-Gaussian fits to the transition
region emission lines of $\alpha$  Cen~A which show broad wings. Flux is in units of  10$^{-15}$.}\label{TtwoG}
 \begin{tabular}{cccc}
 \hline
 \hline
   \multicolumn{4}{l}{\sc Narrow Component} \\
{\sc Ion}&  $v_{rad}$ & FWHM &  Flux\\
  &(km\,s$^{-1}$) & (km\,s$^{-1}$) & (erg s$^{-1}$ cm$^{-2}$)\\
\hline
    Si III 1206.510 &  +5.2$\pm$ 0.4 &48.7$\pm$ 0.4&  1148.2$\pm$ 13.7 \\
 ~~N V   1238.821 &  +5.1$\pm$ 0.8 &42.8$\pm$ 2.4& ~129.5$\pm$  ~9.0 \\
Si IV 1393.755 &  +6.5$\pm$ 0.2 &35.7$\pm$ 0.5&   ~437.9$\pm$  ~6.8 \\
Si IV 1402.770 & +5.4$\pm$ 0.3 &34.8$\pm$ 0.7&   ~221.8$\pm$  ~4.4 \\
C IV 1548.187 &  +8.8$\pm$ 0.2 &43.2$\pm$ 0.4&   ~996.0$\pm$ 11.5 \\
C IV  1550.772 &  +7.8$\pm$ 0.4 &42.3$\pm$ 1.5&   ~478.2$\pm$ 22.9 \\
\hline
Flux-weighted  & & & \\
average  &  +6.8$\pm$ 1.5 &43.4$\pm$ 4.7& ... \\
\hline
\hline
   \multicolumn{4}{l}{\sc  Broad Component} \\
 {\sc Ion} & $v_{rad}$ & FWHM &   Flux\\
& (km\,s$^{-1}$) & (km\,s$^{-1}$) & (erg s$^{-1}$ cm$^{-2}$) \\
\hline
    Si III 1206.510& --0.7$\pm$ 0.7 &69.6$\pm$ 0.4&   745.4$\pm$ 13.7 \\
 ~~N V   1238.821&  +0.9$\pm$ 1.2 &70.4$\pm$ 2.4&   187.8$\pm$  ~7.4 \\
Si IV 1393.755 &  +3.8$\pm$ 0.3 &69.1$\pm$ 0.5& 461.1$\pm$ ~6.8 \\
Si IV 1402.770 &  +3.9$\pm$ 0.6 &65.6$\pm$ 0.8&   228.8$\pm$  ~4.4 \\
C IV 1548.187 & +7.4$\pm$ 0.4 &78.8$\pm$ 0.6&   867.8$\pm$ 11.5 \\
C IV  1550.772 & +4.5$\pm$ 0.6 &72.1$\pm$ 1.7&   474.0$\pm$ 32.5 \\
\hline
Flux-weighted  & & & \\
average    & +3.6$\pm$ 3.1 &72.4$\pm$ 4.4& ... \\
\hline
\hline
   \multicolumn{4}{l}{\sc  Comparison} \\
&  {\sc Flux Ratio} &  {\sc Velocity Shift} & $\chi^2_r$\\
& F$_{BC}$/F$_{tot}$ & ($v_{NC}-v_{BC}$) & \\
 \hline
    Si III 1206.510& 0.39$\pm$ 0.01 & +5.9$\pm$0.8 &  1.12\\
 ~~N V   1238.821& 0.59$\pm$ 0.03 &    +4.3$\pm$1.4 &  1.08\\ 
Si IV 1393.755 &  0.51$\pm$ 0.01 &    +2.7$\pm$0.4 &  1.15\\ 
Si IV 1402.770 &  0.51$\pm$ 0.01 &    +1.5$\pm$0.7 & 1.36\\ 
C IV 1548.187 &  0.47$\pm$ 0.01 & +1.4$\pm$0.5 & 1.17\\ 
C IV  1550.772 &  0.50$\pm$ 0.04 & +3.3$\pm$0.7 & 1.30\\ 
\vspace*{-.1cm}\\
 \hline 
Flux-weighted  & & & \\
average  &   0.46$\pm$ 0.05 & +3.4$\pm$1.8 & \\
\hline
\end{tabular}
\end{table}

The flux-weighted average of the FWHMs are $43.4\pm 4.7$, and $72.4\pm 4.4$
km\,s$^{-1}$ for the narrow and broad components, respectively. By comparison,
explosive  events on the \object{Sun} produce transition region lines as broad as
FWHM$\sim  100$ km\,s$^{-1}$ \citep{dbb}.

 \section{Turbulent velocity and velocity shifts with line formation
temperature}  \label{vshift}

Lines of the same ion generally form at nearly the same temperature in a
collisional ionization equilibrium plasma. Therefore, for most ions we use all
the measured lines to derive their mean Doppler shifts and nonthermal widths.
The results, listed in Table~\ref{tab-vr}, have been derived according to the
following procedure. First, we have computed the standard deviation of the
heliospheric velocities measured for all the unblended lines of each ion. Then,
we have selected  the lines whose velocity is different from the mean by less
than 1 standard deviation in order to remove from the analysis lines that might
be altered by unknown blends or have inaccurate wavelengths. With these
selected lines we then computed the mean heliospheric velocity and standard
deviation of the mean, as well as the mean FWHM. For some ions this procedure
was not applied - e.g. in the case of ions for which less than 3 lines have
been measured - as notated in the last column of Table~\ref{tab-vr}. 
\begin{table}
\caption{Doppler shift and nonthermal velocities of chromospheric and
transition region lines measured in the STIS E140H spectrum of \object{$\alpha$~Cen~A}}
\label{tab-vr}
\begin{tabular}{lcccccl}
\hline
\hline
Ion & Log T$_e$ & N$^a$ & Velocity   & Nonthermal 
 & Notes$^c$ \\
    &                  &                 &Shift$^b$  & Velocity & & \\
     &                  &                 &(km\,s$^{-1}$) &(km\,s$^{-1}$) & &
\\
\hline
Si I     & 3.80 &  80 &   0.00 $\pm$ 0.12 &   7.5 $\pm$  0.3 &   \\
N I      & 3.85 &   5 &  -0.12 $\pm$ 0.10 &  14.6 $\pm$  2.6 &   \\
S I      & 3.95 &  37 &   0.59 $\pm$ 0.07 &   7.6 $\pm$  0.4 &   \\
C I      & 4.11 &  82 &   0.61 $\pm$ 0.09 &   9.9 $\pm$  0.3 &   \\
Fe II    & 4.23 &  90 &   1.50 $\pm$ 0.17 &  10.8 $\pm$  0.3 &   \\
Ni II    & 4.25 &  18 &   1.07 $\pm$ 0.34 &  10.9 $\pm$  0.8 &   \\
Si II    & 4.26 &   9 &   1.21 $\pm$ 0.05 &  23.7 $\pm$  2.3 &   \\
O I      & 4.31 &   3 &   1.01 $\pm$ 0.83 &  11.9 $\pm$  3.2 &   \\
S II     & 4.48 &   2 &   2.27 $\pm$ 1.17 &  15.4 $\pm$  0.4 &   \\
C II     & 4.62 &   2 &   1.68 $\pm$ 0.83 &  27.4 $\pm$  1.3 &    1 \\
C III    & 4.75 &   5 &   3.98 $\pm$ 0.64 &  28.7 $\pm$  0.7 &   \\
Si III   & 4.78 &   3 &   6.92 $\pm$ 0.25 &  27.1 $\pm$  1.3 &   \\
S III    & 4.81 &   1 &   5.05 $\pm$ 0.07 &  22.5 $\pm$  0.3 &    1  \\
Si IV    & 4.84 &   2 &   4.99 $\pm$ 1.47 &  25.7 $\pm$  0.5 &    1  \\
O III    & 4.97 &   2 &   4.73 $\pm$ 1.08 &  19.8 $\pm$  7.3 &    1  \\
C IV     & 5.03 &   2 &   7.28 $\pm$ 0.43 &  30.2 $\pm$  1.0 &    1  \\
O IV     & 5.21 &   2 &   7.25 $\pm$ 1.20 &  27.0 $\pm$  0.5 &   \\
N V      & 5.25 &   2 &   3.87 $\pm$ 0.19 &  30.2 $\pm$  1.1 &    1   \\
S V      & 5.26 &   1 &  11.49 $\pm$ 0.93 &  38.9 $\pm$  0.2 &   \\
O V      & 5.37 &   2 &   6.13 $\pm$ 1.25 &  33.1 $\pm$  0.1 &    1   \\
\hline
\multicolumn{7}{l}{\,$^a$Number of lines selected to compute the velocity shift 
 
}\\
\multicolumn{7}{l}{and nonthermal velocity.}\\
\multicolumn{7}{l}{\,$^b$Doppler shift computed assuming as reference the
measured }\\
\multicolumn{7}{l}{mean velocity of \ion{Si}{i} lines, --23.85 $\pm$0.09
km\,s$^{-1}$.}\\
\multicolumn{7}{l}{\,$^c$We use $1$ in this column to flag the ions for which
all }\\
\multicolumn{7}{l}{ the measured lines were used to compute the velocity shift}\\
\multicolumn{7}{l}{and nonthermal velocity.}\\

\end{tabular}
\end{table}

The most probable  nonthermal  
speeds  ($\xi$) listed in Table~\ref{tab-vr} were
computed from the measured FWHM (in km\,s$^{-1}$) by:
\begin{equation}
\left( \frac{FWHM}{c} \right)^2 = 3.08 \times 10^{-11} \left(
\frac{2kT}{m_i} + \xi ^2 \right),  \label{xsi}
\end{equation}

\noindent where  $c$ is the speed of light, $T$ is the line formation
temperature, and  $m_i$ the ion mass.  The Doppler shifts listed in
Table~\ref{tab-vr} are relative to the the mean velocity of the \ion{Si}{i}
lines (--23.85 $\pm$0.09 km\,s$^{-1}$),  which we use as the reference velocity,
since
the \ion{Si}{i} lines are formed deep in the chromosphere at a temperature
close
to 6500 K \citep{Chaeetal98} and are expected to be at the rest velocity of the
photosphere \citep{Samain91}. This velocity is very close to the radial 
velocity of --23.45 km\,s$^{-1}$ obtained from contemporaneous AAT and CORALIE 
measurements \citep{Pour02}.

\begin{figure}[h]
\begin{centering}
 \resizebox{8cm}{7cm}{\includegraphics{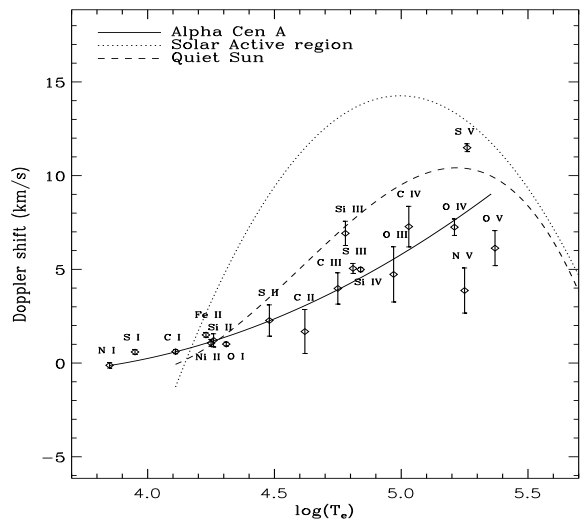}}
 \caption{Doppler shifts of chromospheric and transition region lines of 
\object{$\alpha$~Cen~A} relative to the photospheric radial velocity as a
function of the temperature of line formation. The solid line
represents a fourth order polynomial fit to the data. The dotted and
dashed lines are fits to the Doppler shifts for a solar active region
and the quiet \object{Sun},  respectively,  by \citet{Teriaca99}.}
 \label{figvshift}
\end{centering}
 \end{figure}

\begin{figure}
\begin{centering}
 \resizebox{8cm}{7cm}{\includegraphics{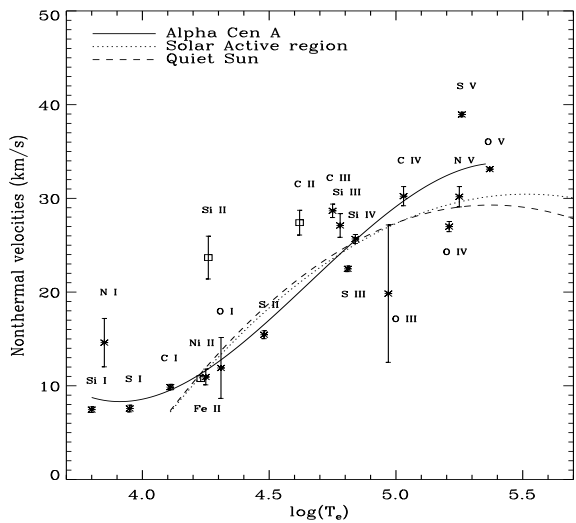}} 
\caption{Nonthermal velocities of chromospheric and transition region 
lines as function of the temperature of line formation.
The solid line represents a third order polynomial fit to the data. 
Lines which are
possibly affected by opacity (\ion{Fe}{ii}, \ion{Si}{ii}, and
\ion{C}{ii}) are not included in the polynomial fit ({\em square data
points}). The dotted and dashed lines represent the nonthermal line
widths in a solar active region and in the quiet \object{Sun},
respectively, as derived by \citet{Teriaca99}.}
 \label{figcsi} 
\end{centering}
 \end{figure}

The Doppler shift and  nonthermal  
widths of the chromospheric and transition
region lines measured in the E140H spectrum of \object{$\alpha$~Cen~A} are plotted in
Figures~\ref{figvshift} and \ref{figcsi}. In both figures polynomial fits to
the {\em SUMER} measurements of radial velocities and nonthermal widths in a
solar active region  and in the quiet \object{Sun}, derived by \citet{Teriaca99}, are
represented with dotted and dashed curves, respectively. According to
\citet{Teriaca99}, in solar active regions the lines formed at temperatures
between T$\sim 2\times 10^4$~K and $\sim 5\times 10^5$~K are red-shifted, with
a maximum red-shift about 15 km\,s$^{-1}$ at $\sim 10^5$~K (\ion{C}{iv}). At
higher
temperatures the velocities decrease becoming blue-shifted (about
--10~km\,s$^{-1}$ at
T$\sim 10^6$~K). However, in the quiet \object{Sun} the Doppler shift reaches a maximum
at a slightly higher temperature,  T$\sim 1.9\times 10^5$~K (\ion{O}{iv},
\ion{N}{v}), and then decreases to a blue-shift  of about --2 km\,s$^{-1}$ at
T$\sim
6.3\times 10^5$~K (\ion{Ne}{viii}). 

We performed a 2nd order polynomial fit to the \object{$\alpha$~Cen~A} line Doppler
shifts with
respect to the temperature of line formation, giving each data point a weight
equal to the square root of the inverse of its standard deviation. We find that
in the sampled temperature range (T$\sim 7\times 10^3$~K to T$\sim 2.3\times
10^5$~K) the redshift increases monotonically, but the data are not adequate to
infer the temperature of the turnover. Even though the lines of \ion{C}{ii},
\ion{Si}{ii}, and \ion{Fe}{ii} are believed to be optically thick, these 
lines do
not show Doppler shifts different from the optically thin lines formed at the
same formation temperature. The \ion{N}{v} lines, especially the 1238~\AA\
line, have broad wings, and their centroids show a smaller redshift than is
expected by the general distribution. In fact, the broad components are
generally marginally blueshifted relative to the narrow components, as
discussed in Section~\ref{BCNC}.
 
For both active and quiet regions on the \object{Sun}, the distributions of nonthermal
line width versus line formation temperature, derived by including only those
lines that are not affected by opacity effects, show a peak at T$\sim 5\times
10^5$~K. To map the increase in turbulent velocity with line formation
temperature (and hence approximate height in the atmosphere), we fit the
\object{$\alpha$~Cen~A} data with a third order polynomial using  the widths of the
optically
thin lines. In the sampled temperature interval, the turbulent velocity
distribution for \object{$\alpha$~Cen~A} resembles the solar data, although slightly
larger
nonthermal line widths are measured for line formation temperatures greater
than T$\sim 10^5$ K. We note that the \ion{O}{iii} mean line width is probably
underestimated because the 1660~\AA\ line may be blended, which can
alter its intensity and consequently the determination of its width.

 \section{Comparison between the UV spectra of \object{$\alpha$ Cen A} and the \object{Sun}}
 \label{sun-comp}
\subsection{The Solar UVSP/SMM  and SUMER/SOHO spectra}

Solar UV spectra with comparable spectral resolution to our HST/STIS spectrum
of \object{$\alpha$~Cen~A} are those observed by the UltraViolet Spectrometer and
Polarimeter
(UVSP) instrument which flew on the the  Solar Maximum Mission (SMM) and by the
SUMER (Solar Ultraviolet Measurements of Emitted Radiation) spectrograph now
operating on SOHO (Solar and Heliospheric Observatory). 

The UVSP/SMM spectrum was obtained  during the minimum of solar cycle (1984) in the 
range 1290 -- 1772 \AA\ with a $1\arcsec \times 180 \arcsec$ slit, oriented north-south  
near solar disk center,
with spectral resolution of the order of 100,000. The atlas (2nd order) was 
prepared
by Richard Shine and Zoe Frank of the Lockheed-Martin Space and Astrophysics 
Lab.,
and was  retrieved  from the site 
\htmladdnormallink{$ftp://umbra.nascom.nasa.gov/pub/uv\_atlases/$}{$ftp://umbra.nascom.nasa.gov/pub/uv\_atlases/$}.
According to Shine (private communication), this spectrum was calibrated using
Rottman's quiet \object{Sun} data from rocket flights, which had accurate flux scales but
had low spectral resolution. For intercomparison purposes, the wavelength scale
of the solar spectrum was shifted by performing a cross correlation between
this spectrum and the \object{$\alpha$ Cen A} spectrum  
in many selected wavelength
intervals. 

The SUMER/SOHO spectrum  is the FUV part of the spectrum that has been derived
from observations obtained  in the range  670 -- 1609~\AA\ by \citet{curdt}.
These data were acquired with a dispersion of  $\sim$41.2 m\AA/pixel (1st
order) at 1500~\AA, for an effective resolution of $\sim$8.2 km\,s$^{-1}$. The
wavelengths are typically accurate to 10 m\AA, i.e. 2 to 5 km s$^{-1}$. The
data represent  the average radiance (mW sr$^{-1}$m$^{-2}$\AA$^{-1}$) for the
quiet \object{Sun} at disk center  (April 20, 1997), a coronal hole  (Oct 12, 1996), 
and a solar spot  (Mar 18, 1999).  Hence, the quiet Sun and coronal hole SUMER spectra 
were acquired during phases of minimum of the solar cycle. 

To be comparable with the \object{$\alpha$~Cen~A} spectrum, we have computed the solar
irradiance at the $\alpha$ Cen distance from the quiet \object{Sun} radiance at disk
center by multiplying by $\pi\ R_{\sun}^2/d_{\alpha Cen}^2$ (cf.
\citealt{Wilhelm98}) for the quiet \object{Sun}, sunspot, and coronal hole spectra. This
conversion does not take into account center-to-limb variations in the lines
and continuum.

\subsection{Comparison between the \object{$\alpha$ Cen A}/STIS and the \object{Sun}/UVSP
spectra}

The comparison between the STIS \object{$\alpha$ Cen A} and the UVSP solar spectra can
be made in the 1192--1688~\AA\ spectral range. In Figure~\ref{uvsp-stis} we
plot interesting regions of the UVSP spectrum and the \object{$\alpha$~Cen~A}  spectrum 
degraded to a resolution of  0.010 \AA/pixel in order to be comparable to the
UVSP spectrum. The wavelength scale of the \object{$\alpha$~Cen~A} spectrum was shifted
to  compensate for  the radial velocity of the system (--23.45 km\,s$^{-1}$). 

\begin{figure}
 \resizebox{8cm}{7cm}{\includegraphics{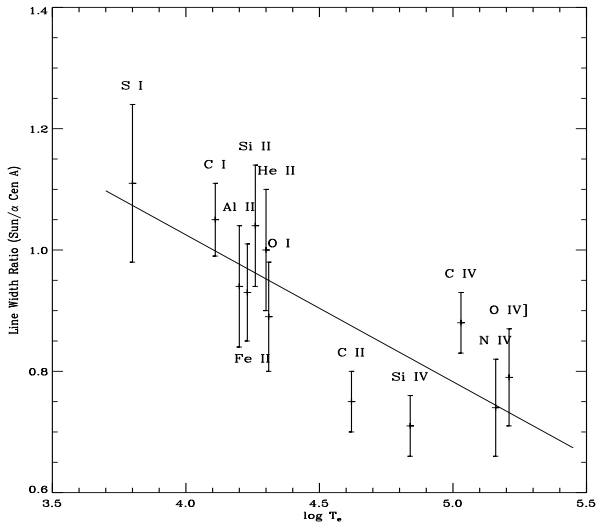}} 
\caption{Ratios of the solar to $\alpha$~Cen~A line widths versus
the temperatures of line formation.}
\label{fwhm}
\end{figure}

\begin{figure}
 \resizebox{8cm}{7cm}{\includegraphics{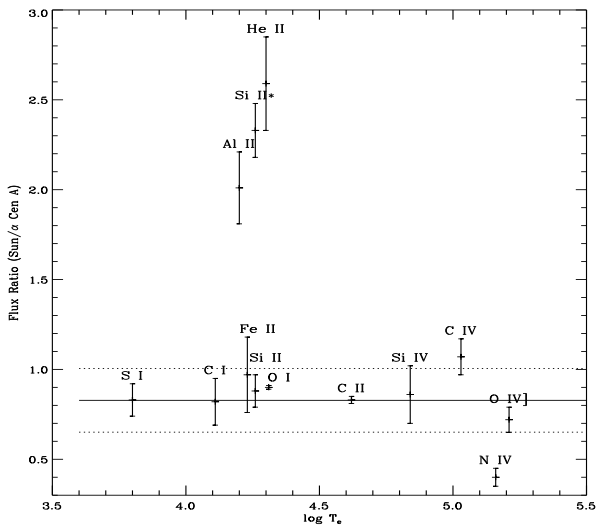}} 
\caption{Ratios of the solar to $\alpha$~Cen~A line surface fluxes 
versus
the temperatures of line formation. The $*$ label  identifies 
the \ion{Si}{ii}  lines at 1526 
and 1533 \AA.  }
\label{surfaceflux}
\end{figure}

Since the UVSP data refer to the ``mean intensity over the disk'', it is
possible to perform a radiometric comparison with the \object{$\alpha$ Cen A} spectrum.
For the emission lines whose
integrated flux in the STIS spectrum exceed 5$\times$ 10$^{-14}$ erg s$^{-1}$
cm$^{-2}$, 
we list  in  Table~\ref{surfacefluxes}  line surface fluxes  
and full widths at  half maximum (FWHM)  for both \object{$\alpha$~Cen~A} 
and the \object{Sun}. We find that the line widths for  
the two stars are very similar for 
most of the
chromospheric lines, whereas the transition region lines are typically broader
for \object{$\alpha$ Cen A} compared to the \object{Sun}. We show in Figure~\ref{fwhm} the FWHM
ratios versus the temperatures of line formation. A linear fit to these data
suggests that the two quantities are correlated with a correlation coefficient
of 0.83.  Typically the \object{$\alpha$ Cen A} line  surface  
fluxes are slightly larger than 
those of the
\object{Sun} (see Figure~\ref{surfaceflux}), with a mean flux ratio 
(\object{Sun}/\object{$\alpha$~Cen~A}) of 0.83$\pm$0.18, but the \ion{Si}{ii} 1526 and 1533 \AA,
the  
\ion{He}{ii}
1640~\AA\ and the \ion{Al}{ii}  1671~\AA\ lines are stronger in the \object{Sun} than in
\object{$\alpha$ Cen A}. The interstellar medium absorption in  the \ion{Si}{ii} 1526  
\AA\ and  \ion{Al}{ii} lines of \object{$\alpha$ Cen A} can  partially explain  the 
high flux ratios. However, this is not the case  for  
the \ion{Si}{ii} 1533  \AA\ 
line. The  common factor  for the three 
lines is the presence of central reversals. 
 In the \object{Sun} these lines form in the
chromosphere, where temperature increases with height. Line source
functions, however, first increase and then decrease with height
over the line formation region, due to non-LTE effects.  A  central
reversal occurs  for an optically thick line when 
the line core forms above the region where
the source function peaks. The ``horns'' of the observed profile
form roughly where the source function peaks \citep{Mauas}. On the \object{Sun} we see 
that the depth of the central reversal is  a function of  
position on the solar 
surface. For example, the \ion{C}{i} lines of the multiplet at 1560 and 1657  
\AA\ show profiles which have central reversals quite deep at the limb, and
less 
pronounced both above the limb
and towards disk center \citep{RD}.
The UVSP spectrum was acquired near disk center, while the \object{$\alpha$~Cen~A}
spectrum is a 
full disk average. As a consequence,  we expect less pronounced central 
reversals for the strongest chromospheric lines 
in the  solar UVSP spectrum than in 
the \object{$\alpha$~Cen~A} spectrum, as  is  observed.
\begin{table*}[ht]
\caption{Surface fluxes (in units of 10$^{3}$
   erg\,cm$^{-2}$\,s$^{-1}$), and FWHM (km\,s$^{-1}$) of a selection of lines present in the spectra
of both  \object{$\alpha$~Cen~A}  and the \object{Sun}.  Included lines
have fluxes in the \object{$\alpha$~Cen~A} spectrum greater than
5$\times$10$^{-14}$ erg s$^{-1}$ cm$^{-2}$. }
\label{surfacefluxes}
\begin{tabular}{cc|ccc|ccc||cc|ccc|ccc}
\hline
\hline
 \multicolumn{1}{c} {Line} &  \multicolumn{1}{c} {Lab}  &
\multicolumn{2}{|c}{Surface Flux} & \multicolumn{1}{c}{Flux}
&\multicolumn{2}{|c}{FWHM}  & FWHM & \multicolumn{1}{|c} {Line} &  \multicolumn{1}{c} {Lab}  &
\multicolumn{2}{|c}{Surface Flux} & \multicolumn{1}{c}{Flux}
&\multicolumn{2}{|c}{FWHM}  & \multicolumn{1}{c}{FWHM}  \\

  \multicolumn{1}{c}{ID} &\multicolumn{1}{c|} {Wavel.} &  \multicolumn{2}{l}{$\alpha$ Cen A}     &\multicolumn{1}{c|}{Ratio} &
 \multicolumn{2}{l}{$\alpha$ Cen} & Ratio &  \multicolumn{1}{|c}{ID} &\multicolumn{1}{c|} {Wavel.} &  \multicolumn{2}{l}{$\alpha$ Cen A}  &  \multicolumn{1}{c|}{Ratio} &
 \multicolumn{2}{l}{$\alpha$ Cen A} & \multicolumn{1}{c}{Ratio}\\

 & \multicolumn{1}{c|}{(\AA )} &\multicolumn{2}{r}{Sun} &
 \multicolumn{1}{c|}{$\frac{Sun}{\alpha Cen A}$}
 & \multicolumn{2}{r}{Sun}
 &$\frac{Sun}{\alpha Cen A}$ &   & \multicolumn{1}{c|}{(\AA )}&\multicolumn{2}{r}{Sun} & 
 \multicolumn{1}{c|}{$\frac{Sun}{\alpha Cen A}$}
 &\multicolumn{2}{r}{Sun}
 &\multicolumn{1}{c}{$\frac{Sun}{\alpha Cen A}$}\\
\hline
         S I & 1300.907 &    0.15 &    0.12 & 0.82 &  13 &  18 & 1.36  & C I+ & 1608.438 &    0.30 &    0.46 & 1.55 &  31 &  28 & 0.92 \\
         O I & 1302.169 &    1.98 &    1.80 & 0.91 &  33 &  32 & 0.96  &	 +Fe II  & & & & & & & \\
       Si II & 1304.372 &    0.18 &    0.16 & 0.88 &  31 &  33 & 1.04   &	       Fe II & 1610.921 &    0.31 &    0.24 & 0.76 &  23 &  20 & 0.89  \\
         O I & 1304.858 &    2.10 &    1.84 & 0.88 &  33 &  27 & 0.80  &	       Fe II & 1611.201 &    0.21 &    0.17 & 0.82 &  17 &  15 & 0.85  \\
         O I & 1306.029 &    2.17 &    1.96 & 0.90 &  29 &  24 & 0.83   &	       Fe II & 1612.802 &    0.39 &    0.34 & 0.87 &  26 &  26 & 0.97 \\
        C II & 1334.532 &    3.27 &    2.68 & 0.82 &  43 &  31 & 0.74   &	         C I & 1613.376 &    0.18 &    0.15 & 0.85 &  12 &  13 & 1.12 \\
        C II & 1335.708 &    4.53 &    3.82 & 0.84 &  47 &  36 & 0.76   &	         C I & 1613.803 &    0.17 &    0.13 & 0.77 &  12 &  13 & 1.07  \\
        Cl I & 1351.657 &    0.30 &    0.32 & 1.06 &  12 &  12 & 0.97 &	         C I & 1614.507 &    0.19 &    0.13 & 0.68 &  13 &  12 & 0.97 \\
         O I & 1355.598 &    0.46 &    0.42 & 0.91 &  12 &  12 & 0.99  &	       Fe II & 1618.470 &    0.25 &    0.30 & 1.21 &  24 &  20 & 0.83 \\
       Si IV & 1393.755 &    2.22 &    2.14 & 0.97 &  44 &  31 & 0.72  &	       Fe II & 1623.091 &    0.28 &    0.25 & 0.89 &  24 &  22 & 0.92  \\
       O IV] & 1401.157 &    0.23 &    0.16 & 0.72 &  46 &  36 & 0.79&	       Fe II & 1625.520 &    0.47 &    0.32 & 0.69 &  27 &  23 & 0.86 \\
       Si IV & 1402.770 &    1.12 &    0.83 & 0.75 &  42 &  29 & 0.69   &	       Fe II & 1625.909 &    0.20 &    0.20 & 0.98 &  17 &  18 & 1.02 \\
      O IV]+ & 1404.806 &    0.12 &    0.08 & 0.66 &  47 &  45 & 0.97   &	       Fe II & 1632.668 &    0.37 &    0.39 & 1.06 &  20 &  18 & 0.90  \\
      +S IV  & & & & & & & &	       Fe II & 1633.908 &    0.37 &    0.32 & 0.86 &  26 &  24 & 0.94 \\
         S I & 1472.972 &    0.38 &    0.28 & 0.74 &  18 &  18 & 1.02   &	       Fe II & 1637.397 &    0.49 &    0.43 & 0.89 &  25 &  24 & 0.94 \\
         S I & 1473.995 &    0.21 &    0.20 & 0.92 &  16 &  15 & 0.96   &	       Fe II & 1640.152 &    0.52 &    0.69 & 1.33 &  22 &  22 & 1.00 \\
        N IV & 1486.496 &    0.16 &    0.06 & 0.40 &  45 &  33 & 0.74    &	       He II & 1640.400 &    0.62 &    1.60 & 2.59 &  52 &  52 & 1.00  \\
       Si II & 1526.708 &    0.67 &    1.64 & 2.44 &  32 &  32 & 0.98    &	       Fe II & 1643.576 &    0.45 &    0.42 & 0.94 &  22 &  22 & 1.00 \\
       Si II & 1533.432 &    0.74 &    1.65 & 2.23 &  37 &  31 & 0.86    &	       Fe II & 1649.423 &    0.30 &    0.21 & 0.71 &  22 &  18 & 0.84 \\
        C IV & 1548.187 &    4.50 &    5.11 & 1.13 &  49 &  45 & 0.88     &	         C I & 1656.260 &    1.68 &    2.05 & 1.22 &  37 &  30 & 0.84 \\
       Fe II & 1550.260 &    0.13 &    0.11 & 0.89 &  20 &  20 & 1.01 &	         C I & 1656.928 &    1.34 &    2.25 & 1.69 &  58 &  54 & 0.92 \\
        C IV & 1550.772 &    2.39 &    2.39 & 1.00 &  52 &  47 & 0.88    &	         C I & 1657.380 &    1.51 &    1.86 & 1.23 &  35 &  31 & 0.89 \\
       Fe II & 1559.084 &    0.47 &    0.39 & 0.82 &  30 &  29 & 0.96    &	         C I & 1657.900 &    0.99 &    1.52 & 1.53 &  28 &  27 & 0.96  \\
          C I & 1560.310 &    0.51 &    0.60 & 1.17 &  27 &  29 & 1.06   &	         C I & 1658.120 &    1.22 &    1.76 & 1.44 &  30 &  26 & 0.87  \\
          C I & 1560.683 &    0.60 &    0.73 & 1.22 &  30 &  36 & 1.21  &	       Fe II & 1658.771 &    0.50 &    0.42 & 0.83 &  24 &  20 & 0.85 \\
          C I & 1561.341 &    0.83 &    1.19 & 1.42 &  56 &  51 & 0.92   &	       Fe II & 1659.483 &    0.75 &    0.69 & 0.91 &  27 &  26 & 0.97 \\
       Fe II & 1563.788 &    0.42 &    0.37 & 0.88 &  28 &  26 & 0.92  &	       O III & 1660.803 &    0.17 &    0.12 & 0.73 &  20 &  22 & 1.07 \\
       Fe II & 1566.819 &    0.30 &    0.25 & 0.81 &  25 &  24 & 0.96  &	       O III & 1666.153 &    0.31 &    0.13 & 0.43 &  45 &  34 & 0.76 \\
       Fe II & 1569.674 &    0.24 &    0.29 & 1.24 &  22 &  22 & 0.99 &	       Fe II & 1669.663 &    0.22 &    0.15 & 0.67 &  16 &  12 & 0.75 \\
       Fe II & 1570.242 &    0.35 &    0.39 & 1.12 &  23 &  25 & 1.07  &	       Al II & 1670.787 &    1.78 &    3.57 & 2.01 &  44 &  41 & 0.94 \\
       Fe II & 1577.166 &    0.15 &    0.19 & 1.22 &  18 &  18 & 0.95 &	       Fe II & 1674.254 &    0.40 &    0.35 & 0.87 &  20 &  17 & 0.85 \\
       Fe II & 1580.625 &    0.30 &    0.30 & 1.02 &  22 &  24 & 1.07  &	      Fe II+ & 1685.954 &    0.29 &    0.38 & 1.30 &  23 &  21 & 0.92  \\
       Fe II & 1584.949 &    0.28 &    0.30 & 1.08 &  23 &  22 & 0.94 &	 +Ni II  & & & & & & & \\
       Fe II & 1588.286 &    0.36 &    0.55 & 1.53 &  21 &  19 & 0.90 &	       Fe II & 1686.455 &    0.44 &    0.55 & 1.26 &  23 &  20 & 0.84  \\
         C I & 1602.972 &    0.13 &    0.13 & 0.99 &  15 &  16 & 1.07 &	       Fe II & 1686.692 &    0.60 &    0.59 & 0.99 &  25 &  26 & 1.04  \\
\hline
\end{tabular}
\end{table*}

 We believe that the \ion{He}{ii} line, which is optically thin and not 
self-reversed,  is really weaker on \object{$\alpha$ Cen A} 
than on the \object{Sun}. Since
the \ion{He}{ii} line is extremely sensitive to the coronal activity, a flux
ratio of 1.8 suggests that the \object{Sun} is more active than \object{$\alpha$ Cen A}. This
conclusion is strengthened by the absence of a limb contribution in the solar
data  since  
the \ion{He}{ii} line is limb-brightened for the \object{Sun} and likely also for
\object{$\alpha$ Cen A}.

  It should be mentioned here that  some of the UV line flux
differences between the  \object{$\alpha$ Cen A} and the \object{Sun} may be due to 
observing at different phases of the two stellar activity cycles. In fact transition 
region lines on the Sun  can vary up to factors 2 -- 5  over a magnetic cycle. While 
both UVSP and SUMER spectra were acquired during phases of solar minimum, we do not 
know the phase of \object{$\alpha$ Cen A} activity cycle at which our observations 
have been obtained. In fact, the long-time extended IUE data base of \object{$\alpha$ Cen A} 
does not provide any hints of an activity cycle period \citep{Ayresetal}. The roughly 
13 individual measurements of the C~IV multiplet flux
obtained by IUE over a 13 year time period (see Fig. 11a in their paper)
have a mean value of $(2.500\pm 0.071) \times 10^{-12}$ but a range from
1.5 to 3.6 in these units. The STIS flux for the C~IV multiplet is 2.80
in the same units. This  STIS flux is about 12\% larger than the mean
IUE flux. Since the IUE data are low resolution (about 6~\AA), the broad
line wings could be difficult to measure compared to the continuum and
nearby weak lines. Thus the C~IV fluxes measured from IUE spectra are
likely somewhat low, and the C~IV flux observed by STIS is probably very
close to the mean value observed by IUE. We therefore believe that
$\alpha$~Cen~A had average transition region fluxes when it was observed
by STIS.

\subsection{Comparison between the \object{$\alpha$ Cen A}/STIS  and \object{Sun}/SUMER spectra}

\begin{figure*}
\begin{centering}
\resizebox{8.0cm}{5cm}{\includegraphics[angle=90]{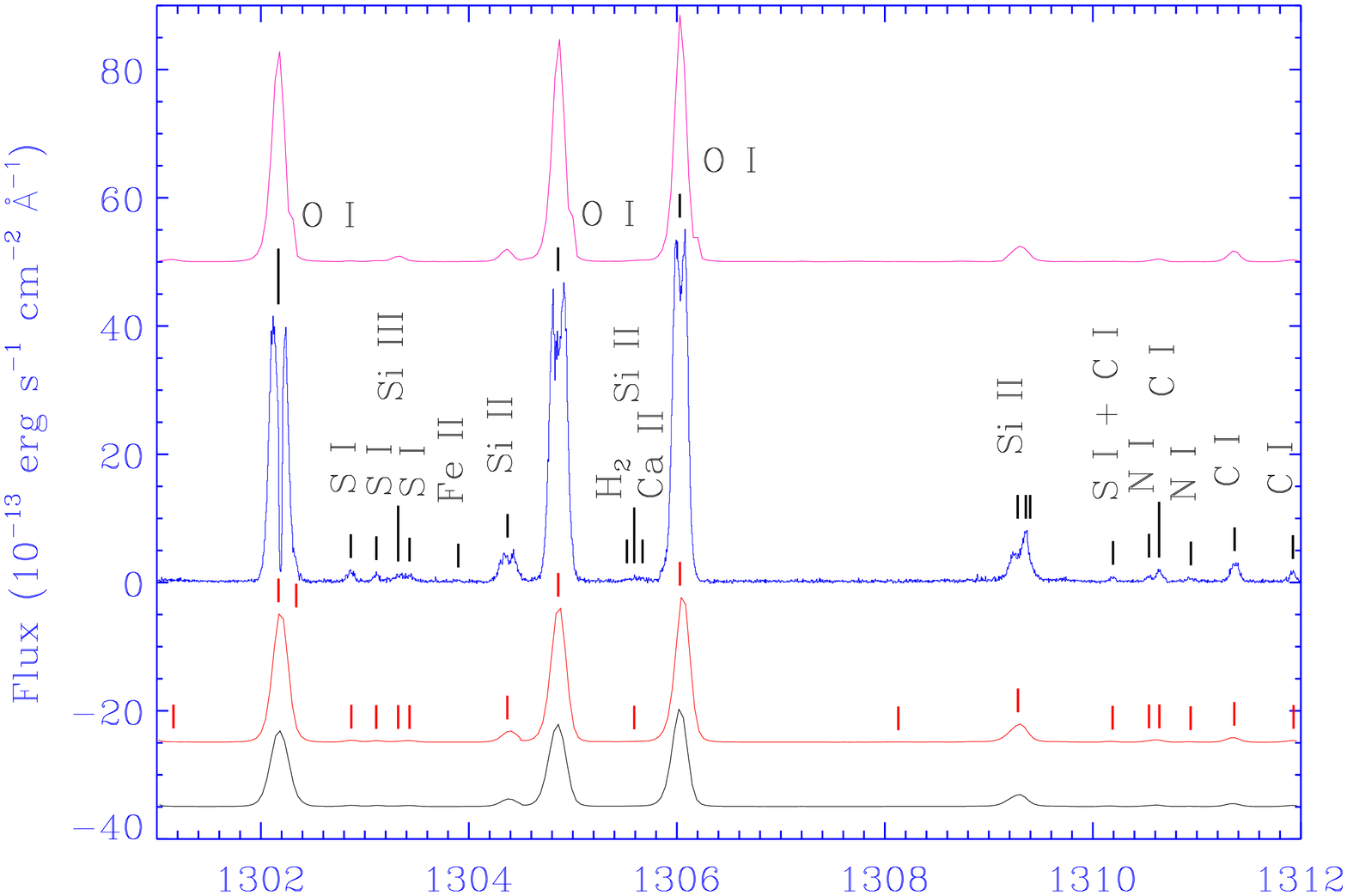}}
\resizebox{8.0cm}{5cm}{\includegraphics[angle=90]{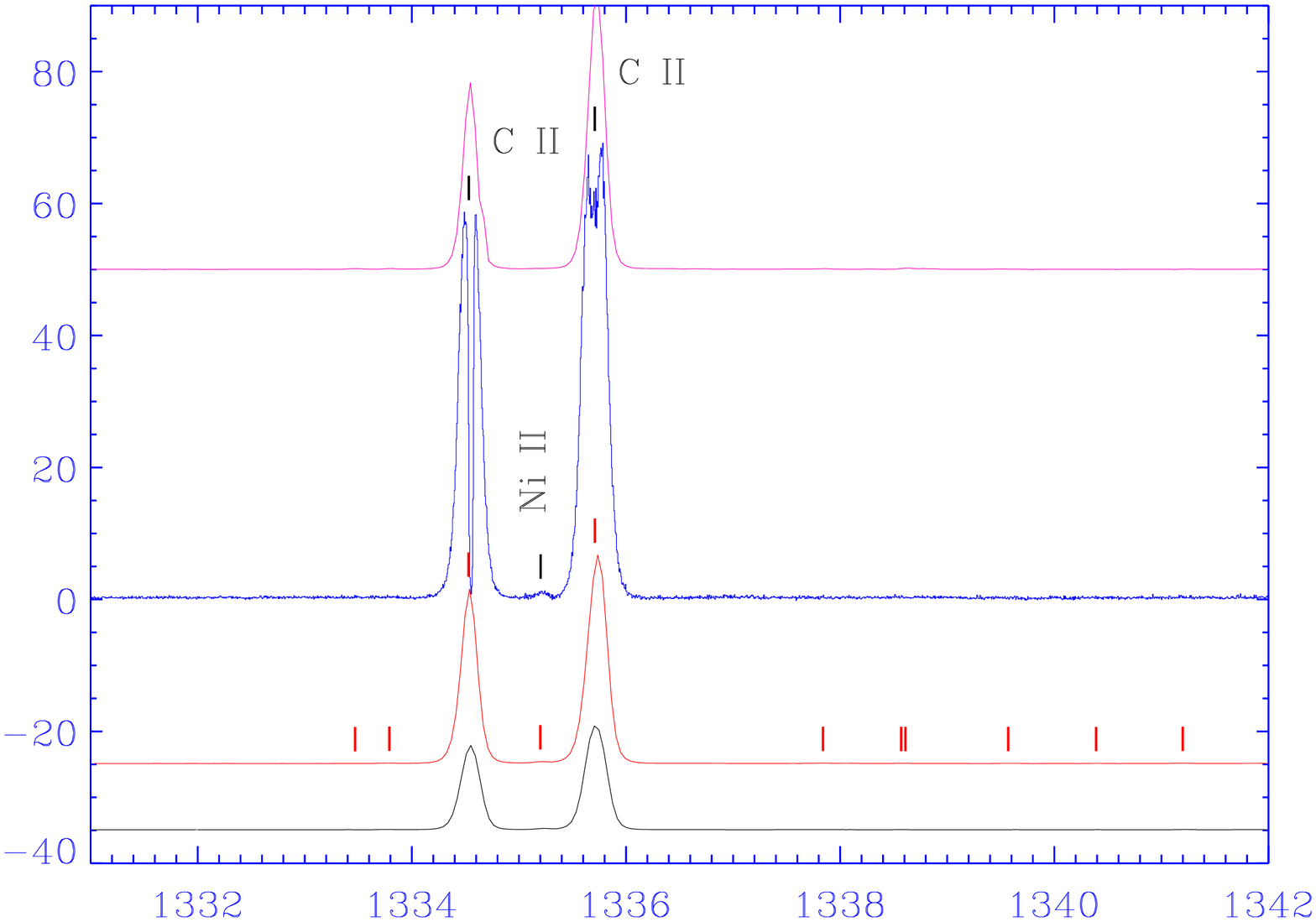}}\\
\resizebox{8.0cm}{5cm}{\includegraphics[angle=90]{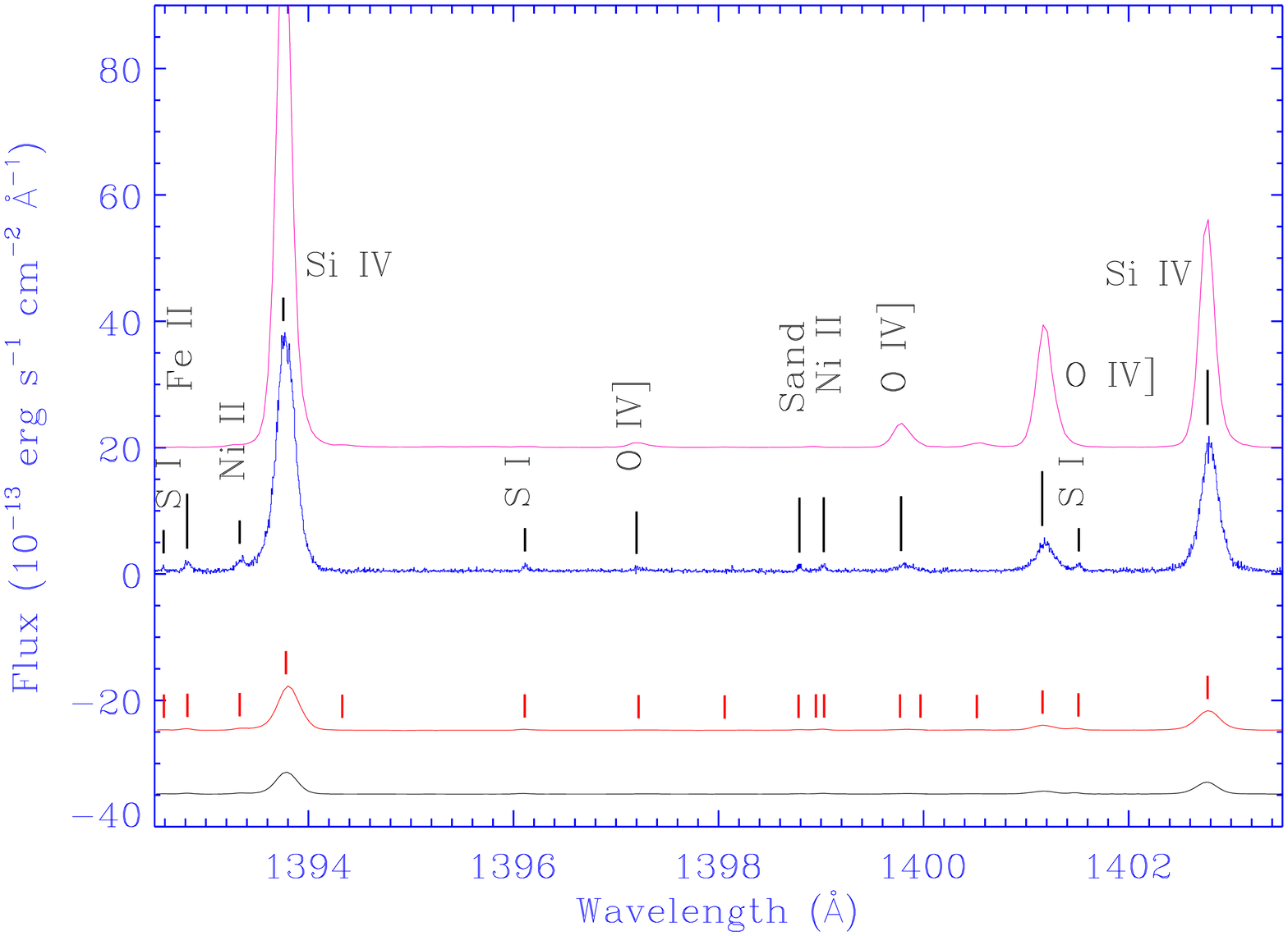}}
\resizebox{8.0cm}{5cm}{\includegraphics[angle=90]{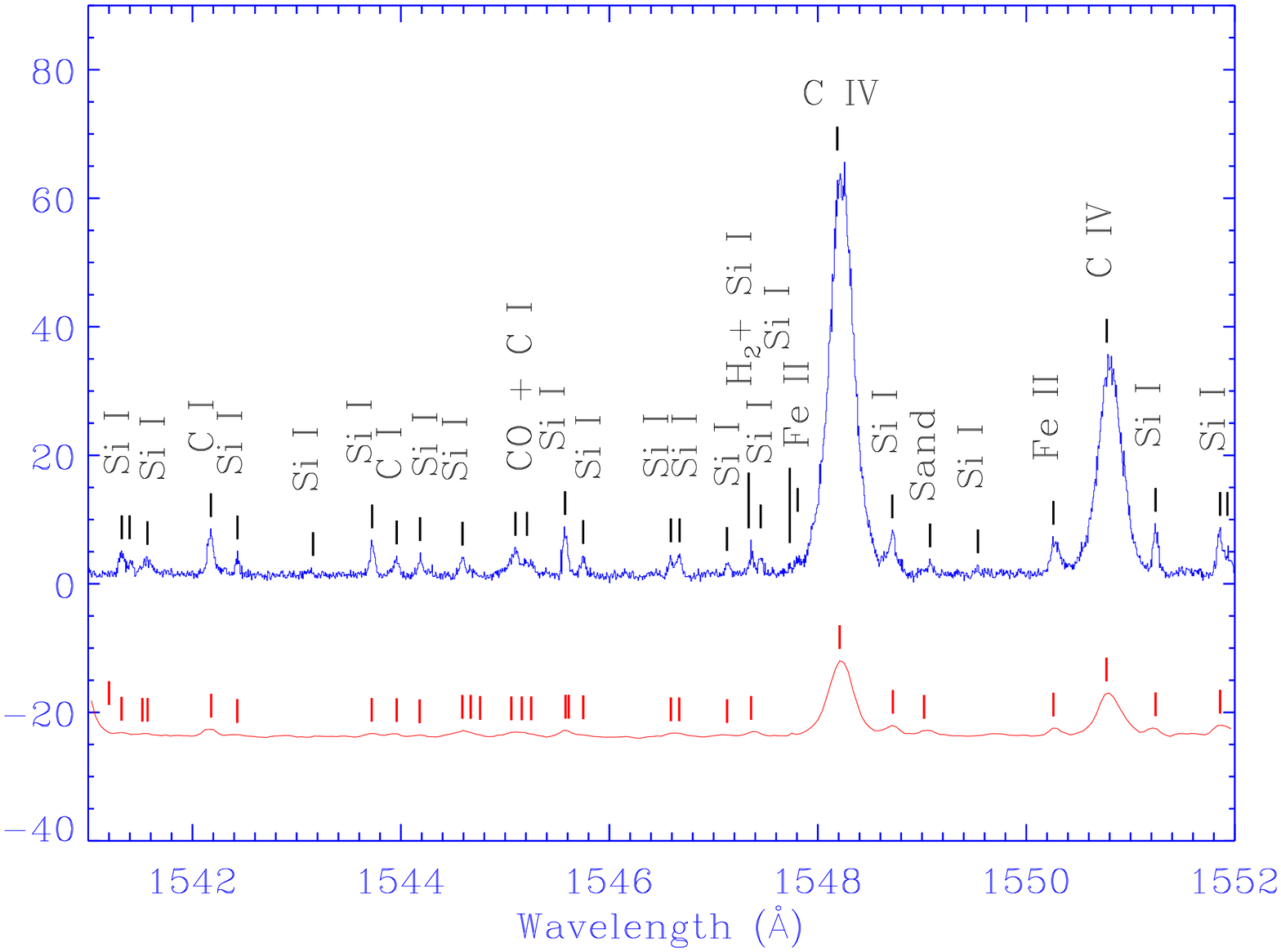}}\\
\caption{Plots of interesting portions of the \object{$\alpha$ Cen A}  HST/STIS 
  and   SOHO/SUMER  quiet sun, coronal hole and solar spot spectra, shifted of   $-$25, $-$35, and +50  in units of the y-axis, respectively (except for the bottom-left panel where the solar spot spectrum is shifted of +20 units). The line identified in the $\alpha$ Cen A STSI spectrum  are labelled, while those identified on the Sun by \citet{curdt}  are marked with vertical lines above the quiet Sun spectrum.}
\label{sumer-stis}
\end{centering}
\end{figure*}

In Figure~\ref{sumer-stis} we plot interesting regions of the \object{$\alpha$~Cen~A}
spectrum,
 and  
the SUMER spectra of a coronal hole, a sunspot, and the quiet \object{Sun},
respectively.  The wavelength scale of the \object{$\alpha$~Cen~A} spectrum was shifted
 to remove the radial velocity of the star (--23.45 km\,s$^{-1}$).  
The SUMER spectrum
has the best photon statistics, therefore faint lines can be more easily seen
in the solar spectrum than in the $\alpha$ Cen STIS spectrum. On the other
hand, the STIS spectrum has better resolution, which can be useful in resolving
line blends and in studying line reversals due to optical thickness effects
better than with the solar spectrum. In Table~\ref{compsumer} we summarize how
many lines we found in common between the two spectra. Many of the lines
present in the solar spectrum but not in \object{$\alpha$ Cen A} are located at
wavelengths below about 1500~\AA, whereas many of the lines detected only in
the STIS spectrum are at wavelengths above 1500~\AA. These differences probably
result from the different resolutions of the two data sets and the increasing
S/N of the SUMER data to shorter wavelengths. The emission lines in the
\object{$\alpha$ Cen A} spectrum are much stronger than in the quiet \object{Sun} spectrum. This
is most likely because we are comparing the solar spectrum, which is an average
disk-center quiet \object{Sun} spectrum, with the \object{$\alpha$ Cen A} full disk irradiance
spectrum, which includes emission from the limb. For most of the lines, the
full disk irradiance is nearly a factor of two larger than the irradiance
derived from disk-center radiance data (cf., \citealt{Wilhelm99}), but there is
no difference for the continuum. Because of this effect, the radiometric
comparison between the SUMER and STIS spectra is uncertain. 

\begin{table}
\caption{Measured features in the 1170-1610~\AA\ common spectral range.}
\label{compsumer}
\begin{tabular}{lccc}
\hline
\hline
 \multicolumn{2}{r}{Total} & 
\multicolumn{1}{c}{With}& \multicolumn{1}{c}{Without}\\
  \multicolumn{2}{r}{Features} & \multicolumn{1}{c}{ID$^a$}
 & \multicolumn{1}{c}{ID$^a$}\\ \hline
 \object{$\alpha$ Cen A}/STIS & 559 & 498 & 61\\
 - of which not in the  solar  spectrum & 172& 127& 45\\
\object{Sun}/SUMER & 516 & 458& 58\\
 - of which not in the \object{$\alpha$ Cen A} spectrum & 126& 80& 46\\
\hline
Lines common to  both spectra & 390 & 377 & 13\\
\hline
\multicolumn{4}{l}{~$^a$ ID = Identification}\\
\end{tabular}
\end{table}

 \section{Electron Densities}
 \label{density}

The  ratios of lines emitted by the same ion can be sensitive to electron
density when the upper levels of the two transitions are depopulated in
different ways. However, misleading results can be obtained when using line
ratios that have a very small sensitivity at the inferred electron densities or
when temperature effects are not properly taken into account. For this reason
we have computed transition region densities for \object{$\alpha$~Cen~A} using both the
line
ratios method and the so-called L-functions method, as described by
\citet{Landi-Landini97}. The main advantage of this method is that it gives an
overall view of all lines and clearly shows which lines (and not line ratios)
are more suitable in a particular density regime and when lines are at the
limit of their density sensitive regime. According to these authors, the {\em
contribution function} for each line of a selected ion, $G_{ij}(T,N_e$), can be
expressed as a product of two functions, one depending on electron density and
electron temperature, and the other on temperature alone: 
\begin{equation}
G_{ij}(T,N_e)= f_{ij}(N_e, T) g(T) .
\label{eqLfun}
\end{equation}

While the $g(T)$ function is mainly determined by the ionization equilibrium
and is the same for all the lines of the same ion, $f(Ne, T)$ is determined
mainly by the population of the upper level. The L-functions are ratios between
the measured intensity of each line of a selected ion and the emissivity
$G_{ij}(T,N_e)$, computed at the temperature $T_{eff}$ where the bulk of the
emission arises. As described by \citet{Landi-Landini97}, the L-functions of
density dependent lines cross each other when plotted versus $\log N_e$ at the
density, or the range of densities, of the region  where the emission arises.
Instead,  the  
L-functions of density independent lines do overlap without 
crossing. The advantages of this method are discussed by \citet{dlm02}. 

The analysis of the \object{$\alpha$~Cen~A} electron densities have been carried out
with the
help of the CHIANTI database VERSION 4.0 (\citealt{dereetal97, Youngetal03}),
assuming the ionization equilibria described by \citet{Mazzottaetal98} and the
\object{$\alpha$~Cen~A} photospheric abundances listed in Table~\ref{abundance}. 


\subsection{\ion{O}{iv} and \ion{S}{iv}}  

The \ion{O}{iv} intercombination multiplet near 1400~\AA\ can be used as a
density diagnostic in the range 10$^9<$N$_e$$<$10$^{12}$ cm$^{-3}$
\citep{bjb96}. The 5 lines of the multiplet are all measured in the
\object{$\alpha$~Cen~A} STIS spectrum. The 1399.780 and 1407.382~\AA\ lines originate
from a
common upper level; their ratio is $1.00 \pm 0.26$, consistent with the ratio
of their A-values (1.08), which is the expected value for a branching ratio in
the optical thin case \citep{Jordan67}. The \ion{O}{iv} 1404.806~\AA\ line is
blended with \ion{S}{iv} 1404.808~\AA, and possibly with another unknown line
\citep{dlm02}.  The percentage of the blending attributable  to the \ion{O}{iv}
line has been derived and discussed by many authors (cf. \citealt{du82, 
bjb96, dlm02}). The analysis of the five \ion{O}{iv} line ratios
(1401/1399, 1401/1407, 1401/1404, 1404/1407, and 1404/1399) indicate that 
$\log N_e$ is in the range 9.8--10.2, assuming that the effective
temperature of \ion{O}{iv} formation is $\log T_e = 5.2$ and that the
\ion{O}{iv}
1404~\AA\ line accounts for about 70-80\% of the blend  
with the \ion{S}{iv}
1404~\AA\ line. This result does not change strongly with the assumed
temperature at which the \ion{O}{iv} lines are formed. Different 
 estimates  of
the \ion{O}{iv} and \ion{S}{iv} relative contributions to the 1404~\AA\ blend
have been found to result in densities inconsistent with those obtained by
ratios not involving the 1404~\AA\ line. This is true if one assumes that
either the \ion{O}{iv} accounts for $\sim$92\% of the blend, which is obtained
from the theoretical line ratio of the \ion{S}{iv} 1404.808~\AA\ and
1406.016~\AA\ lines assuming the atomic calculations by \citet{du82}, or that
the \ion{O}{iv} line accounts for $\sim$50\% of the blend, as derived by
\citet{dlm02} in their analysis of a solar flare and of a GHRS spectrum of
\object{Capella}. On the other hand, in their analysis of FUSE and STIS data for the
dM1e star \object{AU Mic} \citet{dlm02} conclude that the \ion{O}{iv} contribution to
the
blend is $\sim$80\%, which is similar to our analysis.

\begin{figure}
\begin{centering}
\resizebox{8cm}{6cm}{\includegraphics{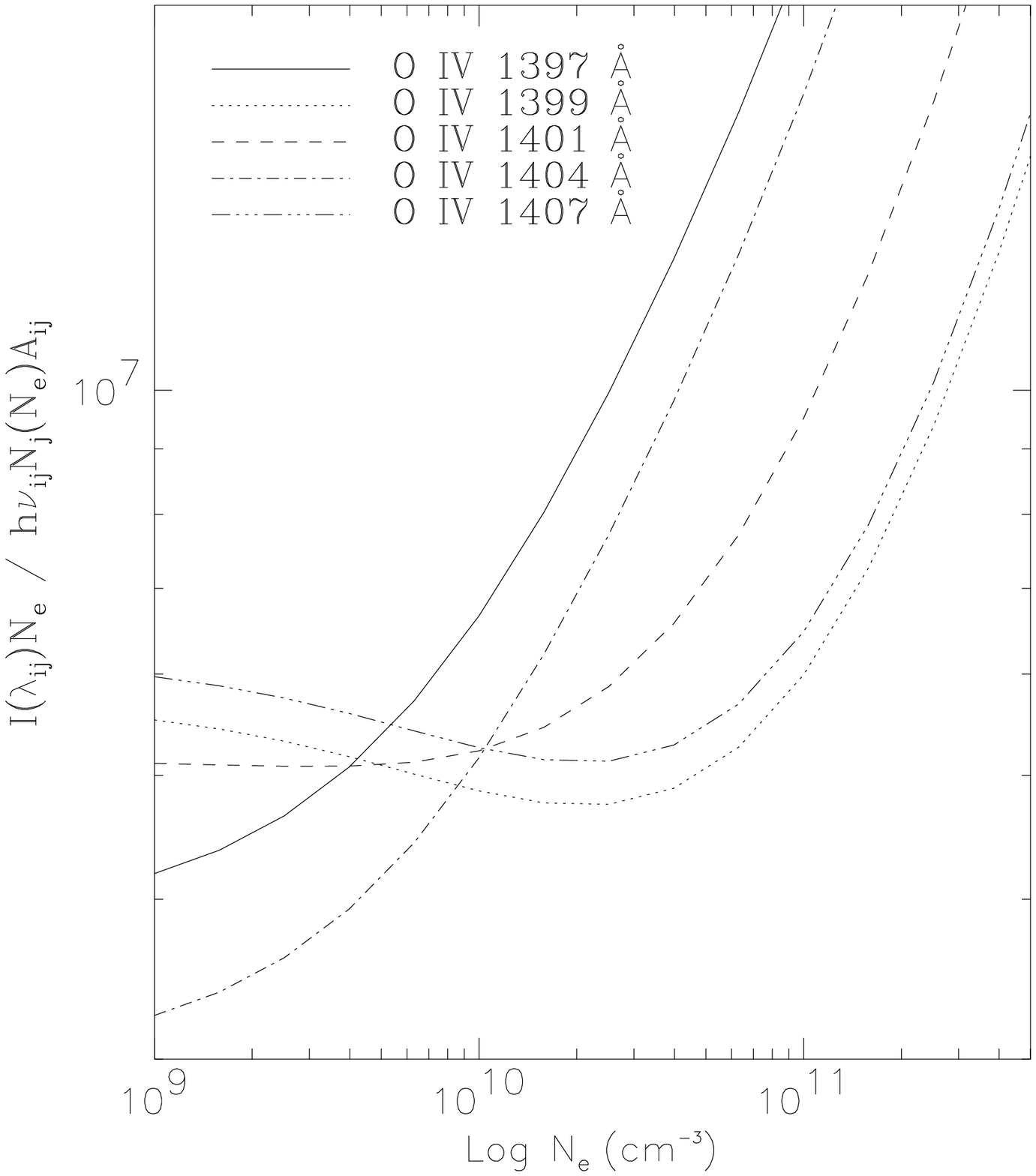}}
\resizebox{8cm}{6cm}{\includegraphics{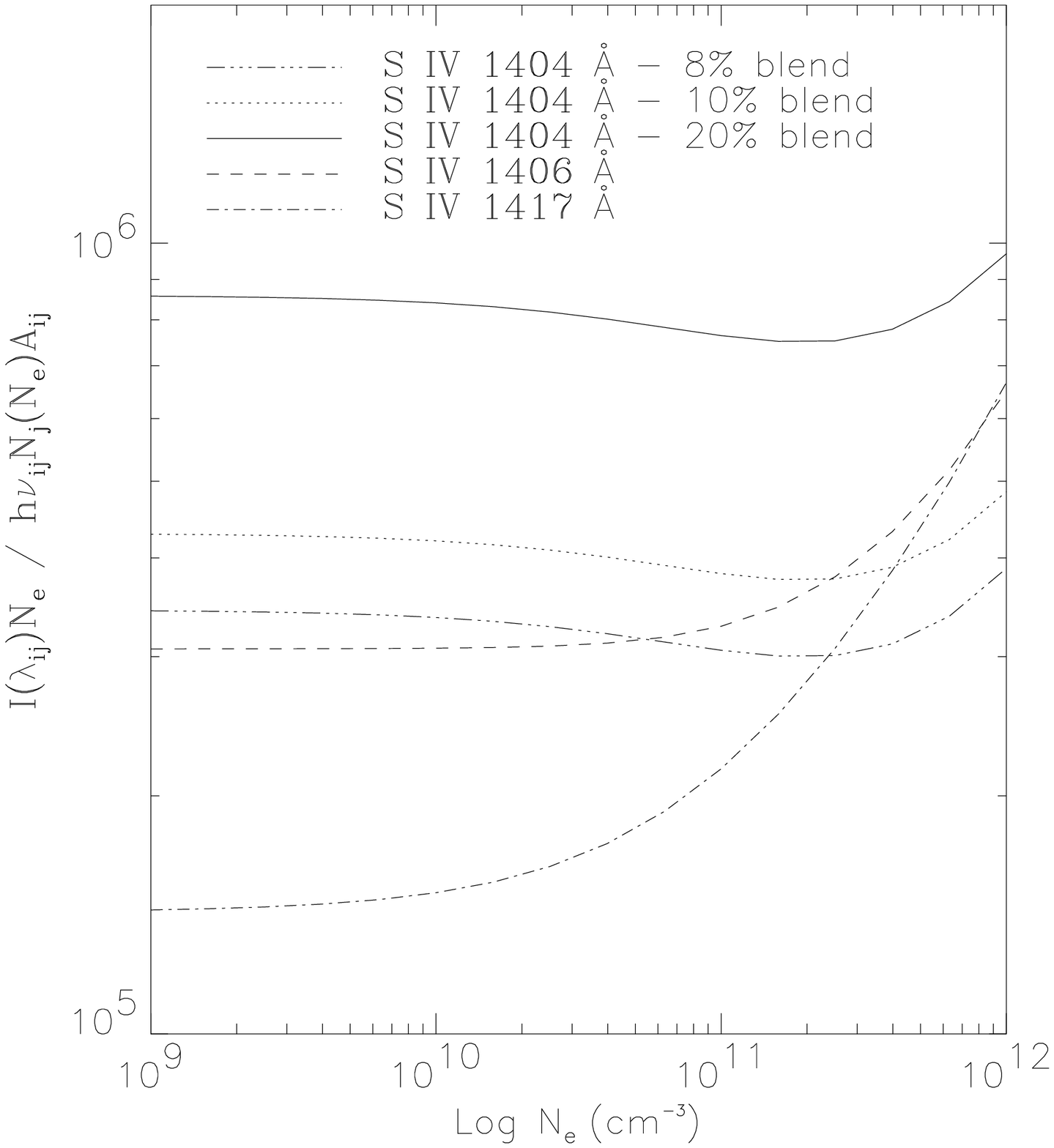}}
\resizebox{8cm}{6cm}{\includegraphics{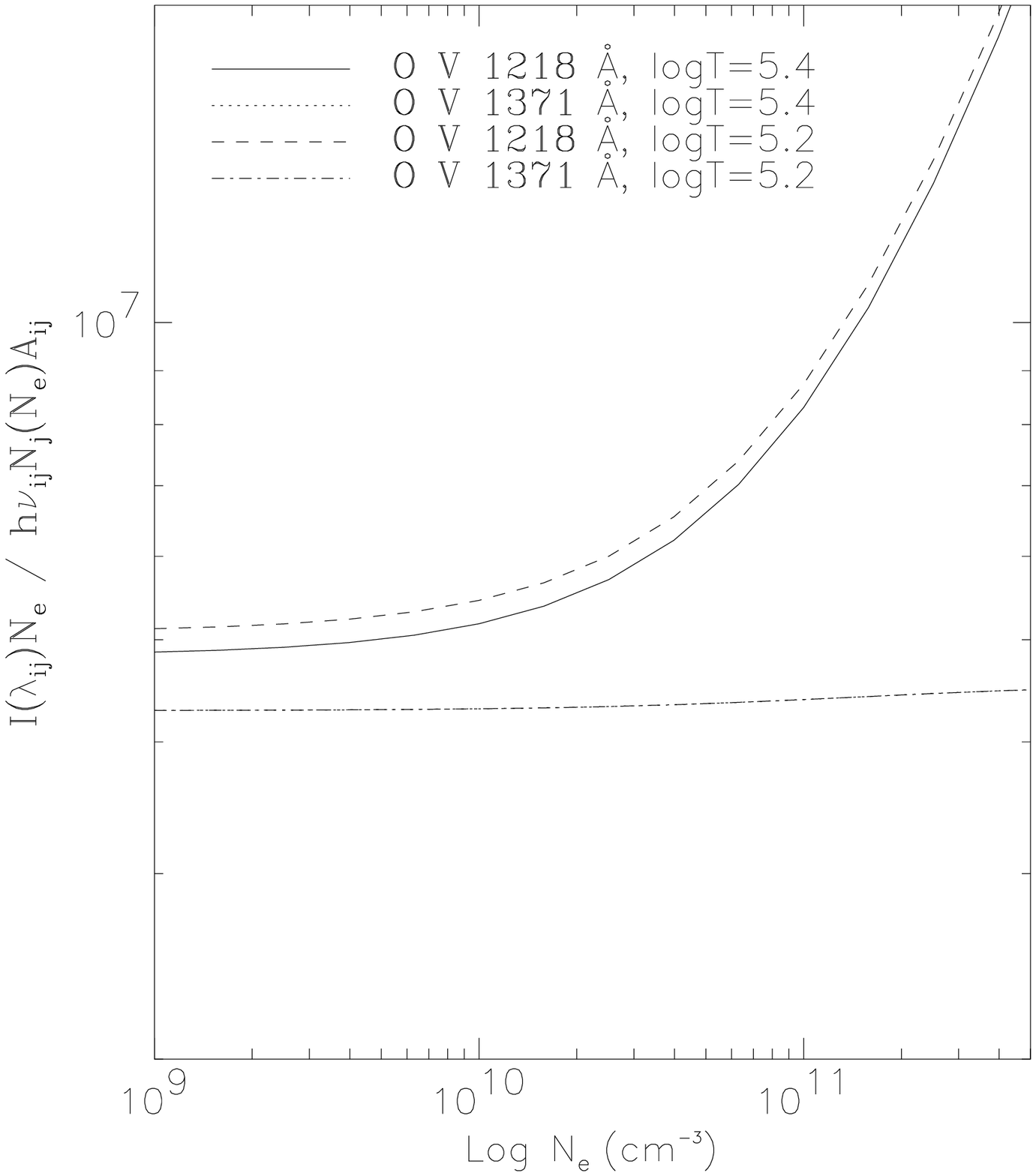}}
\caption{The L-function curves (as defined by Landi \& Landini 1998) plotted for
the \ion{O}{iv}, \ion{S}{iv}, and \ion{O}{v} 1218 and 1371 \AA\  lines
observed in the STIS spectrum of \object{$\alpha$~Cen~A}.}
\label{oivlcurve}
\end{centering}
\end{figure}

In Figure~\ref{oivlcurve} ({\em  top  panel}) 
the L-functions of the \ion{O}{iv}
lines, computed at  $T_{eff} =5.18$, are plotted versus $\log N_e$. Apart from
the  1397~\AA\ line, which is very weak and results   in large errors,
the other lines meet at $\log N_e\sim 10.0$, assuming that the \ion{O}{iv}
1404~\AA\ line accounts for 70\% of the blend. 

The L-functions of the \ion{S}{iv} 1404, 1406, and 1407~\AA\ lines, plotted in
Figure~\ref{oivlcurve} ({\em  middle  panel}), 
show that in order to have
consistency among the lines, the \ion{S}{iv} 1404~\AA\ line must not exceed
10\%
of the flux in the blend. Therefore, an unidentified line contributes $\sim
10-20$\% of the total flux to the 1404~\AA\ blend. Moreover,  the 
L-functions of the
\ion{S}{iv} lines clearly show that no reliable density measurements can be
derived from these lines.

\subsection{\ion{O}{v}} 
\label{densityo5}

The observed \ion{O}{v} 1218/1371 \AA\ line ratio is R=$12.41\pm 2.50$. Using
the Chianti code,  and  
assuming the ionization equilibrium as described in
\citet{Mazzottaetal98} and the \object{$\alpha$~Cen~A} photospheric abundances listed in
Table~\ref{abundance}, we find that: 
\begin{itemize}
\item at the \ion{O}{v} effective temperature, $\log T_e =5.53$, 
the  theoretical 1218/1371 line  ratio varies
in the range $\sim[1,6]$, corresponding to $\log N_e\sim[12,10]$; 

\item at the temperature of \ion{O}{v} maximum ionization fraction, 
$\log T_e =5.40$, the  theoretical line  ratios fall in the
range $\sim [1,7]$, corresponding to $\log N_e\sim[12,9.5]$ .
\end{itemize}

\noindent Therefore, the observed ratio $12.41\pm 2.50$ is not consistent with
the theoretical ratios for any  sensible 
density. In fact, the L-functions of the
\ion{O}{v} 1218 and 1371~\AA\ lines shown in Figure~\ref{oivlcurve} 
 ({\em bottom panel})  at
temperature  $\log T_e =5.4$ do not cross at any density. 

Either  an overestimation of the 1218 \AA\ \ion{O}{v} line flux or an 
underestimation of the 1317 \AA\ line flux can produce  this higher than 
expected flux ratio. The 1218~\AA\ line would have to be less than half the 
measured value, or the 1371~\AA\ line would
have to be more than double the measured value, to be consistent with a density
$\log N_e > 10$ at $\log T_e =5.4$. As discussed in 
Section~\ref{comm} and shown in Figure~\ref{nuovo}, the \ion{O}{v} line at 
1371~\AA\ appears in the STIS
spectrum with a double peak, indicating an apparent central reversal or
overlying absorption. We know of no explanation for this effect, but it could
be the cause of a slight flux understimation for this line.  On the other hand, 
 in Section~\ref{emissionmeasure} we show that the density sensitive \ion{O}{v} 
1218 \AA\ line does not match the differential emission measure distribution
determined from 
the allowed lines at any density, unless we assume that its actual flux is from 
20 to 30\% less than measured.  We think that such an error can be ascribed to   
the difficulty in measuring the \ion{O}{v}  line in the sloping wing of the 
Ly$\alpha$ line, as already discussed  in Section~\ref{comm}, where we  also 
show a slight discrepancy in the  radial velocities measured for the 
\ion{O}{v} 
1218 and 1371 \AA\  lines.

\section{The \object{$\alpha$ Cen A} emission measure distribution}
\label{emissionmeasure}

The physical properties of the transition region plasma can be determined by 
means of
 an emission measure distribution analysis (cf. \citealt{jb81, DeM81, 
pagano00}). 
The frequency integrated flux $F_{ji}$ of an effectively thin emission line
between levels $j$ and $i$ of an atom, in units erg cm$^{-2}$ s$^{-1}$, can be 
written as:
\begin{equation}
\label{fij}
F_{ji}=\frac{hv_{ji}}{2} \int_{\Delta z}\frac{N_jA_{ji}}{N_e^2}\, N_e^2 dz ,
\end{equation}
where $N_j$ (cm$^{-3}$) is the number density of the upper level, $A_{ji}$
(s$^{-1}$) is the Einstein A-coefficient of the line for a transition between
levels $j$ and $i$, and $N_e$ (cm$^{-3}$) is the electron density.  With a high
degree of accuracy, the ratio $\frac{N_jA_{ji}}{N_e^2}$ is a function only of
temperature $T_e$ and density $N_e$. For resonance lines formed at transition
region temperatures, this ratio is a weak function of $N_e$ and a strongly
peaked function of $T_e$, say $f(T_e)$. Therefore, defining the {\em
differential emission measure} $\xi (T_e)$ as: 
\begin{equation}
\xi (T_e)= N_e^2 \frac{dz}{dlog_{10}T_e} ,
\end{equation}
equation~\ref{fij} can be written as:
\begin{equation}
\label{fij1}
F_{ji}=\frac{hv_{ji}}{2} \int_{\Delta log_{10}T_e}f(T_e) \xi(T_e) 
dlog_{10}T_e .
\end{equation}
The differential emission measure $\xi(T_e)$  can in principle be determined
from a set of emission line fluxes, by an inversion of the set of the
corresponding integral equations (eq.~\ref{fij1}). The total emission measure
for the emitting plasma is the integral over $\log T_e$ of the $\xi(T_e)$
function. Since emission lines are typically formed over a temperature range of
$\Delta logT_e\sim 0.3$, we determine the differential emission measure
$\xi^{0.3}(logT_e)$ for all plasma with temperature in the range log\,$T_e
-$0.15 to  log\,$T_e +$0.15.  The shape of the emission measure distribution is
constrained by the combination of the loci from different emission lines, where
the {\em emission measure locus} of each line, in units of cm$^{-5}$, is given 
by:
\begin {equation}
\label{emji}
E_m^{ji}(logT_e)=\int_{\Delta z}N_e^2 dz=\frac{F_{ji}}{\frac{hv_{ji}}{2} 
f(logT_e)}=\frac{F_{ji}}{G_{ji}(logT_e)},
\end{equation} where $E_m^{ji}(logT_e)$ represents the amount of
isothermal material needed to produce the observed line flux.

\begin{figure}
\begin{centering}
 \includegraphics*[angle=90, width=9cm]{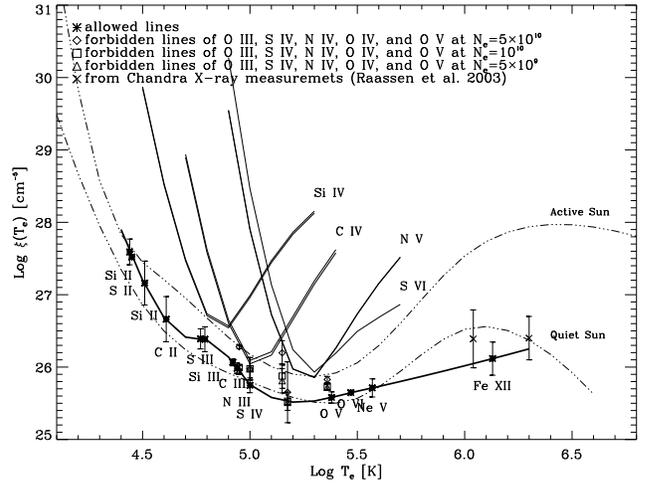}
\caption{The differential  
emission measure distribution of \object{$\alpha$~Cen~A} (solid thick line) is
compared with corresponding distributions for the quiet \object{Sun} and active \object{Sun}. The
intersystem lines of \ion{O}{iii} 1666~\AA, and of the O IV UV 0.01 multiplet
near 1400~\AA\ match the  differential  
emission measure distribution for electron density of
$\log N_e$=9.5--10 [cm$^{-3}$]. The Chandra X-ray data are from
\citet{raassen}.} 
\label{emimes}
 \end{centering}
\end{figure}

We have derived the
emission measure loci as a function of temperature for each emission line of
interest
using equation~\ref{emji}, the $G_{ji}(logT_e)$ functions computed with the
CHIANTI database VERSION 4.0 (\citealt{dereetal97, Youngetal03}), 
the ionization equilibrium as in \citet{Mazzottaetal98}, and the \object{$\alpha$~Cen~A}
photospheric abundances listed in Table~\ref{abundance}. 
Following the procedure described in \citet{pagano00}, we derive the
differential 
emission  measure distribution shown in Figure~\ref{emimes}. We used  the
allowed lines of \ion{Si}{ii}, \ion{S}{ii}, \ion{C}{ii}, \ion{S}{iii}, \ion{Si}{iii},
\ion{C}{iii}, \ion{S}{iv}, \ion{O}{v},  and \ion{Fe}{xii} observed in the STIS
spectrum, which are labelled in the last column of Table~\ref{stis} with the
letters ``{\em emd}''. We have also used the allowed lines \ion{C}{ii} 1036 \&
1037~\AA, \ion{N}{iii} 990~\AA, \ion{S}{iii} 1077~\AA, \ion{Si}{iii} 1108~\AA,
\ion{S}{iv} 1062 \& 1063~\AA, \ion{Ne}{v} 1145~\AA, and \ion{O}{vi} 1037~\AA\
observed in the FUSE spectrum of \object{$\alpha$~Cen~A} \citep{Redfield02}. When more
than
two lines for a given ion have been used, the errorbars in Figure~\ref{emimes}
represent the standard deviation of the emission measures computed for the
different lines. Otherwise, the errorbar is indicative of the uncertainty due
to the line flux measurement. The allowed \ion{C}{iv} 1548 \& 1502~\AA,
\ion{Si}{iv} 1393 \& 1402~\AA, \ion{N}{v} 1238 \& 1242~\AA, and \ion{S}{vi}
933~\AA\ lines were not used to derive the differential emission measure
distribution,
because \citet{dlm02} showed that such lines from the Li and Na isoelectronic
sequences, which were commonly used in previous literature, produce erroneous
results in the determination of emission measures. In Figure~\ref{emimes}
emission measure loci of the \ion{C}{iv}, \ion{Si}{iv}, \ion{N}{v}, and
\ion{S}{vi} lines are, in fact, anomalous with respect to the emission measure
distribution derived from the other ions. 

In Figure~\ref{emimes} we also plot for comparison the differential emission
measure
distribution for the quiet \object{Sun} \citep{Landi-Landini98} and for solar active
regions 
\citep{DeM93}. 
 There is close
agreement between the differential emission measure distributions of
\object{$\alpha$~Cen~A} and the quiet
\object{Sun} in the $\log T$ range 5.0--5.6. For temperatures below $\log T\sim 5.0$,
the
emission measure is larger for \object{$\alpha$~Cen~A} than for the quiet \object{Sun}. At
temperatures
higher than $\log T\sim 5.6$  we have only the
\ion{Fe}{xii} 1242~\AA\ line in the STIS spectrum,  
therefore it is not possible 
to constrain the
real slope of the emission measure distribution. There is, however, a reasonable
good agreement between the  emission measure of \ion{Fe}{xii} 
and the values derived
at temperatures $\log T=6.04$ and 6.3 from Chandra spectra \citep{raassen}.

\begin{figure}
 \begin{centering}
\resizebox{9cm}{6cm}{\includegraphics{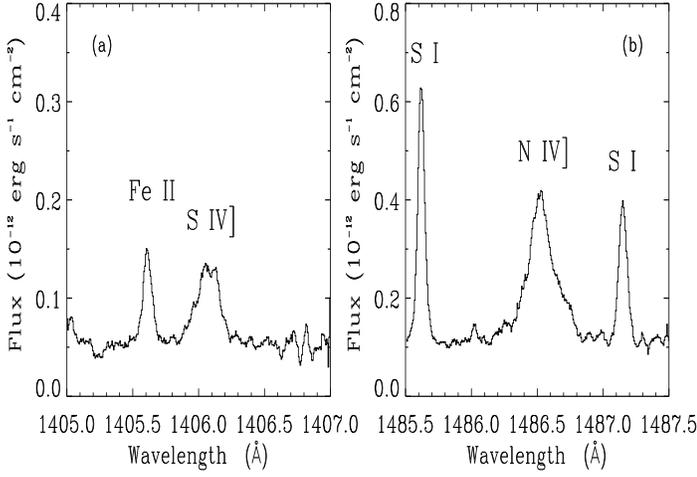}}
\caption{The  S IV] 1406 \AA\ line ({\em panel a}) and the N IV] 1486 \AA\ line 
({\em panel b}) discussed in  Section 7.  } 
\label{s4n4}
 \end{centering}
 \end{figure}

The spin-forbidden lines can be used to obtain information on the plasma
density by comparing their emission measure loci, computed for different values
of the density, with the emission measure distribution derived by using the
resonance lines.  In fact, for collisionally de-exicited spin-forbidden lines
(i.e., when $N_eC_j>A_{ji}$ and 
 $C_j=\sum_iC_{ji}$  cm$^{-3}$ s$^{-1}$ is the
total collision rate out of level $j$), the ratio $N_jA_{ji}/N_e^2$ in
Equation~\ref{fij} is proportional to  $f(T_e)/N_e$. Therefore, the emission
measure loci of spin-forbidden lines depend upon both electron density and
temperature.  In Figure~\ref{emimes} we have plotted the emission measures of
the intersystem \ion{O}{iii} 1666\AA, \ion{S}{iv} 1406~\AA, \ion{N}{iv} 1486~\AA,
and \ion{O}{v} 1218~\AA\ lines, and the mean emission measure of the
\ion{O}{iv}
lines at 1397, 1399, 1401, and 1407~\AA. The \ion{O}{iii} and \ion{O}{iv} ions
match the emission measure distribution derived from allowed lines for electron
density log\,$N_e$=9.5--10, and lie above at higher densities. The \ion{S}{iv},
\ion{N}{iv}, and \ion{O}{v} lines do not strictly match the emission measure
distribution at any density. A possible explanation for this behaviour is that 
the fluxes of these lines are slightly overestimated (no more 
than 50\%). As shown in Figure~\ref{s4n4}, the     \ion{S}{iv} 1406~\AA\ and 
\ion{N}{iv}  1486~\AA\ lines are very well detected in the STIS \object{$\alpha$~Cen~A}
spectrum, 
and a flux overestimation could be caused only by  unknown blends. 
Alternatively,  inaccurate atomic data  could be the cause of the 
observed discrepancy.  For  the \ion{O}{v} 1218~\AA\ line,   a flux 
 from 20 to 30\% less than measured would  make 
this line consistent with the 
emission measure distribution derived from the allowed lines. As seen in  
Section~\ref{densityo5},  the anomalous  ratio  
between 1218 and 1371 \AA\ line 
fluxes also suggests that the  \ion{O}{v} 1218~\AA\ line flux was
overestimated, 
possibly because of the sloping Ly$\alpha$ wings (cf. Section~\ref{comm}).     

The total power radiated per unit (surface) area from the stellar
atmosphere is:
\begin{equation}
P=\int_{T_{\rm min}}^{T_{\rm max}}\frac{N_H}{N_e} P_{\rm rad}(T_e) 
\frac{\xi(T_e)}{T_e} dT_e ,
\label{prad}
\end{equation}

\noindent where $P_{\rm
rad}$ is the total power radiated in all spectral features per unit emission
measure and for a specified set of atomic abundances. Above log\,$T_e\sim 4.0$ 
hydrogen is predominantly ionized, therefore $N_H/N_e\sim 0.8$. By using the
analytical expression for $P_{rad}$ given by \citet{RTV} 
and the differential emission measure derived above, we find a radiative power
loss of  1.5$\times$10$^6$ erg s$^{-1}$ cm$^{-2}$, i.e. 2.4$\times 10^{-5} 
L_{\rm bol}$, for the plasma in the
temperature range log\,$T_e=4.4-6.5$. This result is compatible with
the estimate of the total power radiated between 10$^4$ and 10$^8$~K, excluding
hydrogen emission, which can be  computed from the surface flux of the \ion{C}{iv}
lines by using the \citet{BMCW} empirical relation, P=f(CIV 1548+1550
\AA)/3.0$\times$ 10$^{-4}$. For \object{$\alpha$~Cen~A} this formula predicts
P=5.7$\times
10^{-5} L_{\rm bol}$. 
Focusing on only the  regions for which  we have derived a very  well 
constrained  emission measure distribution, that are those in the temperature 
range  log\,$T_e=4.4-5.6$,  we find that their radiative power loss is  
4.2$\times$10$^5$ erg s$^{-1}$
cm$^{-2}$ or 6.5$\times 10^{-6} L_{\rm bol}$. 

 With the same method described above, we have computed the power loss from 
Solar 
Active regions and from the Quiet \object{Sun} adopting their emission measure 
distributions as in Figure~\ref{emimes}.  For the temperature range 
range log\,$T_e=4.4-5.6$ we find a power loss of 8.3$\times$10$^5$ erg s$^{-1}$
cm$^{-2}$ (1.3$\times 10^{-5} L_{\rm bol}$) from Solar Active Regions, and of 
2.6$\times$10$^5$ erg s$^{-1}$
cm$^{-2}$ (4.0$\times 10^{-6} L_{\rm bol}$) from the Quiet \object{Sun}. The
\object{$\alpha$~Cen~A} 
power loss in the same temperature range is therefore midway between those of the 
Quiet Sun and the Solar Active 
Regions, but closer to the former than  to 
the latter.

 \section{The Interstellar medium  towards $\alpha$ Cen}
 \label{ism-sec}

A number of lines in the \object{$\alpha$~Cen~A} spectrum contain interstellar
absorption 
components (look for label "ISM" in the $Notes$ column of the
Table~\ref{stis}). The analysis of these components will be presented and
analyzed in a subsequent 
paper.
We call here attention to the narrow 
absorption features in the \ion{Si}{iv} 1393 \AA, \ion{C}{iv} 1548 \AA, and
\ion{N}{v} 1238 \AA\  lines shown in Figure~\ref{fig3}.  We think that these 
features could be real. In fact, if they were an artifact of the
line spread function, then they would appear in all emission
lines, not just the high excitation lines.  Also the possibility that these 
features are the results of order 
overlapping, in the case the wavelength
scales of two adjacent orders were not consistent, was examined and
disregarded. 
 In fact, not one of the above emission lines lies in the overlap region 
of two adjacent orders. Although the central reversals in \ion{N}{v}
 and perhaps 
\ion{C}{iv} could be noise, the reversal in the\ion{Si}{iv} line appears to be 
too deep and includes too many pixels to be just noise. The reality of the 
Si IV feature suggests that the other features could be real. If the 
reversals are real, then what is their cause? Self-absorption should 
affect both lines in a doublet, but the weaker members of the doublet do 
not show absorption features. Perhaps we are detecting narrow absorption 
by some cool species above the hot regions. Since these possible 
absorption features are too narrow to be thermal, we have no explanation 
for them, and further observations and analysis are needed to verify 
their reality and, if real, search for their cause.

 \section{Conclusions}
\label{conclu}

We present our analysis of HST/STIS observations of \object{$\alpha$~Cen~A} and compare
its
spectrum with its near twin, the \object{Sun}: 

(1) We present a high resolution ($\lambda/\Delta\lambda = 114,000$) spectrum
of \object{$\alpha$~Cen~A} obtained using the E140H mode of STIS that covers the
spectral range
1140--1670~\AA\ with very high signal-to-noise. The spectrum has an absolute
flux calibration accurate to  $\pm 5$\%, an absolute wavelength accuracy of
0.6--1.3 km\,s$^{-1}$, and and is corrected for scattered light. To our knowlege this 
is the best available ultraviolet spectrum of a solar-like star. 

(2) As strange as this may at first appear, there is no available ultraviolet
reference spectrum of the \object{Sun} as a point source with the characteristics of the
\object{$\alpha$~Cen~A} spectrum that can be used to compare stellar spectra with the
\object{Sun}. Many
ultraviolet spectra of the \object{Sun} do exist, but they either have lower spectral
resolution, lack wavelength or flux accuracy, or do not include the
center-to-limb variation across the solar disk required to provide an accurate
spectrum of the \object{Sun} as a point source. Although \object{$\alpha$~Cen~A} differs slightly
from
the \object{Sun} in effective temperature, gravity, and metal abundance, its spectrum
can serve as a representative solar spectrum for comparison with other stars. 

(3) We compare the \object{$\alpha$~Cen~A} spectrum to the solar irradiance (the \object{Sun}
viewed as
a point source) derived from UVSP data for the ``mean intensity over the disk''
by placing the \object{Sun} at the distance to \object{$\alpha$~Cen~A} and shifting the
\object{$\alpha$~Cen~A} spectrum
by the star's radial velocity. The line widths of the two stars are similar for
chromospheric lines, but the transition region lines of \object{$\alpha$~Cen~A} are
broader
than those of the \object{Sun} by roughly 20\%. The line surface fluxes are typically larger on
\object{$\alpha$~Cen~A}, presumably due to \object{$\alpha$~Cen~A} being somewhat metal rich.
However, the
\ion{He}{ii} 1640~\AA\ line is stronger in the \object{Sun}, indicating that the solar
corona is more active. 

(4) We also compare the \object{$\alpha$~Cen~A} spectrum to the solar irradiance derived
from
SUMER spectra of the disk center quiet \object{Sun}, assuming constant center-to-limb
radiance and shifting the \object{$\alpha$~Cen~A} wavelength scale by the radial
velocity of
the star. A total of   671 emission lines are detected in the \object{$\alpha$~Cen~A}
spectrum
from 37 different ions and 2 molecules (CO and H$_2$). In addition to the well
known chromospheric and transition region lines, we also identify lines of
\ion{Al}{iv}, \ion{Si}{viii}, \ion{S}{v}, \ion{Ca}{vii}, \ion{Fe}{iv}, \ion{Fe}{v},
and \ion{Fe}{xii}. A total of 172 emission lines observed in \object{$\alpha$~Cen~A} are
not
seen in the SUMER spectrum. 

(5) Broad wings are present in the strong  resonance  
lines of \ion{C}{iv}, \ion{N}{v},
\ion{Si}{iii}, and \ion{Si}{iv}, as are seen in solar observations of the
chromospheric network. We fit the line profiles with two Gaussians: a narrow
component ascribed to Alfv\'en waves in small magnetic loops, and a broad
component ascribed to microflares or magneto-acoustic waves in large coronal
funnels. Both components are redshifted with the narrow Gaussians having larger
redshifts as is seen on the \object{Sun}. At line formation temperatures between
20,000~K and 200,000~K, there is a trend of increasing line redshift, similar
to but with a somewhat lower magnitude than the quiet \object{Sun}. A similar trend of
increasing nonthermal velocities with temperature is nearly identical to that
which is observed in solar quiet and active regions. 

(6) Using line ratios and L-functions, we infer that the \ion{O}{iv} lines 
are formed where the electron density is $\log N_e \sim 10.0$. The \ion{S}{iv} 
and \ion{O}{v} lines, however, do not provide reliable $N_e$ values.
 Values of $N_e$ have been obtained for
the Sun and other solar type stars (c.f. \citealt{cook}). 
It is hard to make any comparison with these results because they are strongly affected by   
the adopted atomic calculation or by the choice of lines  with a limited density sensitivity, 
as the \ion{O}{iv} 1400 \AA\ line  (see \citealt{dlm02}). Hence estimates of  $N_e$   from 
different computations are often not consistent. We can, however, compare the electron 
density we derive for 
$\alpha$~Cen~A at $\log T\sim 5.2$ with the electron density derived by \citet{dlm02} for  
\object{Capella} (G1 III + G8 III)  and \object{AU Mic} (dM1e), because we use the same 
computation methods. The comparison tell us that at $\log T \sim 5.2$ the electron density 
is slightly less   in $\alpha$ Cen A  than in the more active Capella ($\log N_e \sim  10.6$) 
and AU~Mic ($\log N_e \sim  10.7$).

(7) The emission measure distribution of \object{$\alpha$~Cen~A} derived from emission
lines of
ions not in the Li and Na isoelectronic sequences is in close agreement with
 that of  
the quiet \object{Sun} in the temperature range $5.0 < \log T < 5.6$, but lies somewhat 
above the quiet \object{Sun} in the temperature range $4.5 < \log T < 5.0$. This
could be explained by the higher metal abundance of \object{$\alpha$~Cen~A} combined
with a
somewhat less active corona that provides less conductive heating to the upper
transition region. The estimated total radiative power loss from the transition
region ($4.4 < \log T < 5.6$) is $4.2\times 10^5$ erg s$^{-1}$ cm$^{-2}$,
corresponding to $2.4\times 10^{-5} L_{\rm bol}$.

 \begin{acknowledgements}
This work is supported by NASA grant S-56500-D to NIST and the University of
Colorado. We thanks Dr. Joseph B. Gurman, who kindly provided us information
about the UVSP/SMM solar spectrum, and Dr. Richard Shine and Dr. Zoe Frank who
made it available. 
\end{acknowledgements}




 \clearpage


\setcounter{table}{3}

\begin{deluxetable}{llrrrrl}
 \tabletypesize{\small}
\tablewidth{0pt}
\tablecaption{Line identifications, velocity shifts, FWHM, and fluxes. \label{Tsp}
} 
\tablehead{
\colhead {Line$^a$} &  \colhead {Lab}  &\colhead {Observed} & \colhead {Velocity$^b$}& \colhead {Flux}&  \colhead {FWHM$^c$}  &\colhead{Notes$^d$} \\ 
 \colhead {ID}&  \colhead {Wavelength} &\colhead {Wavelength} & \colhead{Shift} &
 \colhead {(10$^{-15}$)} & \colhead {} & \\
& \colhead {(\AA)} &    \colhead {(\AA)} &  \colhead{ km\,s$^{-1}$} &   \colhead {erg\,cm$^{-2}$\,s$^{-1}$} &  \colhead {  km\,s$^{-1}$}  & } 
\startdata
          C III  &             1174.933 & 1174.8595 $\pm$ 0.0009 &    4.7 $\pm$ 0.2 &      162.0 $\pm$     3.7 &     47.7 $\pm$ 0.4 &                            emd \\
          C III  &             1175.263 & 1175.1910 $\pm$ 0.0008 &    5.1 $\pm$ 0.2 &      125.2 $\pm$     3.8 &     49.1 $\pm$ 0.5 &                            emd \\
          C III  &             1175.590 & 1175.4908 $\pm$ 0.0070 &   -1.8 $\pm$ 1.8 &      111.2 $\pm$     7.1 &     49.5 $\pm$ 0.4 &                            emd \\
          C III  &             1175.711 & 1175.6449 $\pm$ 0.0031 &    6.6 $\pm$ 0.8 &      402.2 $\pm$     4.7 &     48.4 $\pm$ 0.4 &                            emd \\
          C III  &             1175.987 & 1175.9141 $\pm$ 0.0008 &    4.8 $\pm$ 0.2 &      114.0 $\pm$     3.7 &     48.0 $\pm$ 0.7 &                            emd \\
          C III  &             1176.370 & 1176.2869 $\pm$ 0.0008 &    2.3 $\pm$ 0.2 &      155.8 $\pm$     3.9 &     51.2 $\pm$ 1.8 &                            emd \\
          Si II  &             1190.416 & 1190.3302 $\pm$ 0.0009 &    1.8 $\pm$ 0.2 &       37.7 $\pm$     3.0 &     39.5 $\pm$ 0.5 &                       emd, ISM \\
          Si II  &             1193.289 & 1193.2031 $\pm$ 0.0008 &    1.9 $\pm$ 0.2 &       23.6 $\pm$     4.3 &     56.0 $\pm$ 0.4 &                       emd, ISM \\
\enddata
\tablecomments {~\\
\noindent  $\,^a$ Line identification is according to \citet{san86} unless differently noted in the last column. \\
\noindent  $\,^b$ Radial velocity shifts corrected for the  stellar radial velocity
of --23.45 km\,s$^{-1}$, computed using the orbital parameters and ephemeris given  by
\citet{Pour02};\\
\noindent  $\,^c$ {\em UL} means unresolved line;\\
\noindent $\,^d$ {\em CR} means line with Central Reversal; {\em ISM} means presence 
of an absorption component due to interstellar medium absorption; 
{\em emd} means that the line was used to derive the differential emission 
measure distribution; {\em NC} and {\em BC} means that the line profile can be fit by a
 broad plus a narrow components. 
 Not identified lines present also in the solar spectrum, and
classified by \citet{san86} in four classes, are listed with the
name {\em Sandlin-class\_number}  and the observed wavelentgh in the Solar 
spectrum; {\em NIST} means that the line has been identified according to the NIST
 Atomic Spectra Database, NIST Standard Reference Database \#78 
 (http://physics.nist.gov/cgi-bin/AtData/main\_asd)
 \\ 
\noindent $\,^e$ For the laboratory wavelengths of O IV] and S IV] lines in the 1400 \AA\ region see \citet{Harper}.\\
\noindent $\,^f$ He II multiplet: 1640.332/.345/ .375/.391/.474/.490/.533\\
}
\end{deluxetable}
\normalsize

\end{document}